\documentclass[a4paper,UKenglish,cleveref, autoref, thm-restate]{lipics-v2021}
\nolinenumbers

\title{Optimal Oblivious Algorithms for Multi-way Joins}

\author{Xiao Hu}{University of Waterloo, Canada}{xiaohu@uwaterloo.ca}{https://orcid.org/0000-0002-7890-665X}{}
\author{Zhiang Wu}{University of Waterloo, Canada}{zhiang.wu@uwaterloo.ca}{https://orcid.org/0009-0004-8647-1416}{}
\authorrunning{X. Hu and Z. Wu}

\usepackage{amsthm, amsmath, mathtools}
\usepackage[linesnumbered,vlined,ruled,commentsnumbered]{algorithm2e}
\usepackage{booktabs}
\usepackage{rotating}
\renewcommand{\paragraph}[1]{\smallskip \noindent {\bf #1}}
\newcommand{\OUT}{\textrm{OUT}}

\newcommand{\Q}{\mathcal{Q}}
\newcommand{\V}{\mathcal{V}}
\newcommand{\E}{\mathcal{E}}

\newcommand{\T}{\mathcal{T}}

\newcommand{\dom}{\mathrm{dom}}

\newcommand{\eat}[1]{}

\newcommand{\wcoj}{\textrm{WCOJ}}

\newcommand{\A}{\mathcal{A}}

\newcommand{\val}{\textsf{\upshape val}}
\newcommand{\key}{\textsf{\upshape key}}

\newcommand{\R}{\mathcal{R}}

\newcommand{\re}{\textbf{read}\ }
\newcommand{\wt}{\textbf{write}\ }



\EventEditors{}
\EventNoEds{2}
\EventLongTitle{28th International Conference on Database Theory (ICDT 2025)}
\EventShortTitle{ICDT 2025}
\EventAcronym{ICDT}
\EventYear{2025}
\EventDate{}
\EventLocation{}
\EventLogo{}
\SeriesVolume{}
\ArticleNo{}

\begin{document}

\maketitle

\begin{abstract}
In cloud databases, cloud computation over sensitive data uploaded by clients inevitably causes concern about data security and privacy. 
Even when encryption primitives and trusted computing environments are integrated into query processing to safeguard the actual contents of the data, access patterns of algorithms can still leak private information about the data. {\em Oblivious RAM} (ORAM) and {\em circuits} are two generic approaches to address this issue, ensuring that access patterns of algorithms remain oblivious to the data. However, deploying these methods on insecure algorithms, particularly for multi-way join processing, is computationally expensive and inherently challenging. 

In this paper, we propose a novel sorting-based algorithm for multi-way join processing that operates without relying on ORAM simulations or other security assumptions. Our algorithm is a non-trivial, provably oblivious composition of basic primitives, with time complexity matching the insecure worst-case optimal join algorithm, up to a logarithmic factor. Furthermore, it is {\em cache-agnostic}, with cache complexity matching the insecure lower bound, also up to a logarithmic factor. This clean and straightforward approach has the potential to be extended to other security settings and implemented in practical database systems.
\end{abstract}

\section{Introduction}
\label{sec:intro}
In outsourced query processing, a client entrusts sensitive data to a cloud service provider, such as Amazon, Google, or Microsoft, and subsequently issues queries to the provider. The service provider performs the required computations and returns the results to the client. Since these computations are carried out on remote infrastructure, ensuring the security and privacy of query evaluation is a critical requirement. Specifically, servers must remain oblivious to any information about the underlying data throughout the computation process. To achieve this, advanced cryptographic techniques and trusted computing hardware are employed to prevent servers from inferring the actual contents of the data~\cite{hacigumucs2002executing,costan2016intel}. However, the memory accesses during execution may still lead to information leakage, posing an additional challenge to achieving comprehensive privacy. For example, consider the basic (natural) join operator on two database instances: $R_1 = \left\{(a_i,b_i): i\in [N] \right\} \Join S_1 = \left\{(b_i, c_i): i \in [N]\right\}$ and $R_2 = \left\{(a_i, b_1): i \in [N] \right\} \Join S_2 = \left\{(b_1,c_i): i \in [N] \right\}$ for some $N \in \mathbb{Z}^+$, where each pair of tuples can be joined if and only if they have the same $b$-value. 
Suppose each relation is sorted by their $b$-values.
Using the merge join algorithm, there is only one access to $S_1$ between two consecutive accesses to $R_1$, but there are $N$ accesses to $S_2$ between two consecutive accesses to $R_2$. Hence, the server can distinguish the degree information of join keys of the input data by observing the sequence of memory accesses. Moreover, if the server counts the total number of memory accesses, it can further infer the number of join results of the input data.

The notion of \textit{obliviousness} was proposed to formally capture such a privacy guarantee on the {\em memory access pattern} of algorithms~\cite{goldreich1996software, goldreich1987towards}. This concept has inspired a substantial body of research focused on developing algorithms that achieve obliviousness in practical database systems~\cite{zheng2017opaque, eskandarian2017oblidb, crooks2018obladi, chang2022towards}. A generic approach to achieving obliviousness is {\em Oblivious RAM} (ORAM)~\cite{goldreich1996software, kushilevitz2012security, gentry2013optimizing, wang2015circuit, devadas2016onion, stefanov2018path, asharov2020optorama}, which translates each logical access into a poly-logarithmic (in terms of the data size) number of physical accesses to random locations of the memory. 
but the poly-logarithmic additional cost per memory access is very expensive in practice~\cite{chang2016oblivious}. Another generic approach involves leveraging {\em circuits}~\cite{wang2022query,fan2024tight}. Despite their theoretical promise, generating circuits is inherently complex and resource-intensive, and integrating such constructions into database systems often proves to be inefficient. These challenges highlight the advantages of designing algorithms that are inherently oblivious to the input data, eliminating the need for ORAM frameworks or circuit constructions. 

In this paper, we take on this question for {\em multi-way join processing}, and examine the insecure {\em worst-case optimal join} (\wcoj) algorithm~\cite{ngo2018worst, ngo2014skew,veldhuizen2014leapfrog}, that can compute any join queries in time proportional to the maximum number of join results. Our objective is to investigate the intrinsic properties of the \wcoj~algorithm and transform it into an oblivious version while preserving its optimal complexity guarantee.

\subsection{Problem Definition}
\label{sec:preliminary}
\noindent {\bf Multi-way join.} A (natural) join query can be represented as a hypergraph $\Q = (\V,\E)$~\cite{abiteboul1995foundations}, where $\V$ models the set of attributes, and $\E \subseteq 2^{\V}$ models the set of relations. Let $\dom(x)$ be the domain of attribute $x \in \V$. An instance of $\Q$ is a function $\R$ that maps each $e \in \E$ to a set of tuples $R_e$, where each tuple $t \in R_e$ specifies a value in $\dom(x)$ for each attribute $x \in e$. The result of a join query $\Q$ over an instance $\mathcal{R}$, denoted by $\Q(\R)$, is the set of all combinations of tuples, one from each relation $R_e$, that share the same values in their common attributes, i.e., 
\begin{equation*}
\label{eq:QR}
    \Q(\R) = \left\{t \in \prod_{x \in \V} \dom(x) \mid \forall e \in \E, \exists t_e \in R_e, \pi_{e} t = t_e\right\}.
\end{equation*} 

\noindent Let $N = \sum_{e \in \E} |R_e|$ be the {\em input size} of instance $\R$, i.e., the total number of tuples over all relations. Let $\OUT = |\Q(R)|$ be the {\em output size} of the join query $\Q$ over instance $\R$. We study the data complexity~\cite{abiteboul1995foundations} of join algorithms by measuring their running time in terms of input and output size of the instance. We consider the size of $\Q$, i.e., $|\V|$ and $|\E|$, as constant.

    \paragraph{\bf Model of computation.} We consider a two-level hierarchical memory model \cite{krastnikov2020efficient, chu2021differentially}. The computation is performed within \textit{trusted memory}, which consists of $M$ registers of the same width. For simplicity, we assume that the trusted memory size is $c \cdot M$, where $c$ is a constant. This assumption will not change our results by more than a constant factor. Since we assume the query size as a constant, the arity of each relation is irrelevant. Each tuple is assumed to fit into a single register, with one register allocated per tuple, including those from input relations as well as intermediate results. We further assume that $c\cdot M$ tuples from any set of relations can fit into the trusted memory. Input data and all intermediate results generated during the execution are encrypted and stored in an {\em untrusted memory} of unlimited size. Both trusted and untrusted memory are divided into blocks of size $B$. One memory access moves a block of $B$ consecutive tuples from trusted to untrusted memory or vice versa. The complexity of an algorithm is measured by the number of such memory accesses. 

    An algorithm typically operates by repeating the following three steps: (1) read encrypted data from the untrusted memory into the trusted memory, (2) perform computation inside the trusted memory, and (3) Encrypt necessary data and write them back to the untrusted memory.  Adversaries can only observe the address of the blocks read from or written to the untrusted memory in (1) and (3), but not data contents. They also cannot interfere with the execution of the algorithm. The sequence of memory accesses to the untrusted memory in the execution is referred to as the ``access pattern'' of the algorithm. In this context, we focus on two specific scenarios of interest:
    \begin{itemize}
        \item {\bf Random Access Model (RAM).} This model can simulate the classic RAM model with $M=O(1)$ and $B=1$, where the trusted memory corresponds to $O(1)$ registers and the untrusted memory corresponds to the main memory. The {\em time complexity} in this model is defined as the number of accesses to the main memory by a RAM algorithm.
        \item {\bf External Memory Model (EM).} This model can naturally simulate the classic EM model~\cite{aggarwal1988input,vitter2001external}, where the trusted memory corresponds to the main memory and the untrusted memory corresponds to the disk. Following prior work \cite{frigo1999cache, demaine2002cache, chu2021differentially}, we focus on the {\em cache-agnostic} EM algorithms, which are unaware of the values of $M$ (memory size) and $B$ (block size), a property commonly referred to as {\em cache-oblivious} in the literature. To avoid ambiguity, we use the terms ``cache-agnostic'' to refer to ``cache-oblivious'' and ``oblivious'' to refer to ``access-pattern-oblivious'' throughout this paper. The advantages of cache-agnostic algorithms have been extensively studied, particularly in multi-level memory hierarchies. A cache-agnostic algorithm can seamlessly adapt to operate efficiently between any two adjacent levels of the hierarchy. We adopt the {\em tall cache} assumption, $M = \Omega(B^2)$ and further $M = \Omega(\log^{1+\epsilon} N)$\footnote{In this work, $\log(\cdot)$ always means $\log_2(\cdot)$ and should be distinguished from $\log_{\frac{M}{B}} (\cdot)$.} for an arbitrarily small constant $\epsilon \in (0,1)$, and the {\em wide block} assumption, $B = \Omega(\log^{0.55} N)$. These are standard assumptions widely adopted in the literature of EM algorithms~\cite{aggarwal1988input,vitter2001external, arge2007optimal,frigo1999cache, demaine2002cache, chu2021differentially}. The {\em cache complexity} in this model is defined as the number of accesses to the disk by an EM algorithm.
    \end{itemize}

    \noindent {\bf Oblivious Algorithms.} The notion of obliviousness is defined based on the access pattern of an algorithm. Memory accesses to the trusted memory are invisible to the adversary and, therefore, have no impact on security. Let $\A$ be an algorithm, $\Q$ a join query, and $\R$ an arbitrary input instance of $\Q$.  We denote $\mathsf{Access}_{\A}(\Q,\R)$ as the sequence of memory accesses made by $\mathcal{A}$ to the untrusted memory when given $(\Q,\R)$ as the input, where each memory access is a read or write operation associated with a physical address. The join query $\Q$ and the size $N$ of the input instance are considered non-sensitive information and can be safely exposed to the adversary. In contrast, all input tuples are considered sensitive information and should be hidden from adversaries. This way, the access pattern of an oblivious algorithm $\mathcal{A}$ should only depend on $\Q$ and $N$, ensuring no leakage of sensitive information.

    \begin{definition}[Obliviousness~\cite{goldreich1987towards,goldreich1996software,chan2019foundations}] 
    \label{def:oblivious} 
         An algorithm~$\mathcal{A}$ is oblivious for a join query $\Q$, if given an arbitrary parameter $N \in \mathbb{Z}^+$, for any pair of instances $\R, \R'$ of $\Q$ with input size $N$, $\mathsf{Access}_{\A}(\Q, \R) \overset{\mathrm{\delta}}{\equiv}\mathsf{Access}_{\A}(\Q,\R')$, where $\delta$ is a negligible function in terms of $N$. Specifically, for any positive constant $c$, there exists $N_c$ such that $\delta(N) < \frac{1}{N^c}$ for any $N>N_c$. The notation $\overset{\mathrm{\delta}}{\equiv}$ indicates the statistical distance between two distributions is at most $\delta$. 
 \end{definition}
 
    This notion of obliviousness applies to both deterministic and randomized algorithms. 
    For a randomized algorithm, different execution states may arise from the same input instance due to the algorithm's inherent randomness. Each execution state corresponds to a specific sequence of memory accesses, allowing the access pattern to be modeled as a random variable with an associated probability distribution over the set of all possible access patterns. The statistical distance between two probability distributions is typically quantified using standard metrics, such as the total variation distance. A randomized algorithm is indeed oblivious if its access pattern exhibits statistically indistinguishable distributions across all input instances of the same size. Relatively simpler, a deterministic algorithm is oblivious if it displays an identical access pattern for all input instances of the same size.

    \subsection{Review of Existing Results}
    \noindent {\bf Oblivious RAM.} ORAM is a general randomized framework designed to protect access patterns~\cite{goldreich1996software}. In ORAM, each logical access is translated into a poly-logarithmic number of random physical accesses, thereby incurring a poly-logarithmic overhead. Goldreich et al.~\cite{goldreich1996software} established a lower bound $\Omega(\log N)$ on the access overhead of ORAMs in the RAM model. Subsequently, Asharov et al.~\cite{asharov2020optorama} proposed a theoretically optimal ORAM construction with an overhead of $O(\log N)$ in the RAM model under the assumption of the existence of a one-way function, which is rather impractical~\cite{shi2020path}. It remains unknown whether a better cache complexity than $O(\log N)$ can be shown for such a construction. Path ORAM~\cite{stefanov2018path} is currently the most practical ORAM construction, but it introduces an $O(\log^2 N)$ overhead and requires $\Omega(1)$ trusted memory. In the EM model, one can place the tree data structures for ORAM in an Emde Boas layout, resulting in a memory access overhead of $O(\log N \cdot \log_B N)$. 

   \paragraph{Insecure Join Algorithms.} The \wcoj~algorithm \cite{ngo2018worst} have been developed to compute any join query in $O(N^{\rho^*})$ time\footnote{A hashing-based algorithm achieves $O(N^{\rho^*})$ time in the worst case using the lazy array technique~\cite{flum2002query}.}, where $\rho^*$ is the fractional edge cover number of the join query (formally defined in Section~\ref{sec:agm-preliminary}). The optimality is implied by the AGM bound~\cite{atserias2008size}. \footnote{The maximum number of join results produced by any instance of input size $N$ is $O(N^{\rho^*})$, which is also tight in the sense that there exists some instance of input size $N$ that can produce $\Theta(N^{\rho^*})$ join results.} However, these \wcoj~algorithms are not oblivious. In Section~\ref{sec:triangle}, we use triangle join as an example to illustrate the information leakage from the \wcoj~algorithm. Another line of research also explored output-sensitive join algorithms. A join query can be computed in $O((N^\textsf{subw} + \OUT)\cdot \mathsf{polylog} N))$ time~\cite{yannakakis1981algorithms, abo2016faq}, where $\textsf{subw}$ is the submodular-width of the join query. For example, $\textsf{subw}=1$ if and only if the join query is acyclic~\cite{beeri1983desirability, fagin1983degrees}. 
   These algorithms are also not oblivious due to various potential information leakages. For instance, the total number of memory accesses is influenced by the output size, which can range from a constant to a polynomially large value relative to the input size. A possible mitigation strategy is {\em worst-case padding}, which involves padding dummy accesses to match the worst case. However, this approach does not necessarily result in oblivious algorithms, as their access patterns may still vary significantly across instances with the same input size.
   
    In contrast, there has been significantly less research on multi-way join processing in the EM model. First of all, we note that an EM version of the \wcoj~ algorithm incurs at least $\Omega\left(\frac{N^{\rho^*}}{B}\right)$ cache complexity since there are $\Theta(N^{\rho^*})$ join results in the worst case and all join results should be written back to disk. For the basic two-way join, the nested-loop algorithm has cache complexity $O\left(\frac{N^2}{B}\right)$ and the sort-merge algorithm has cache complexity $O\left(\frac{N}{B} \log_{\frac{M}{B}} \frac{N}{B} + \frac{\OUT}{B}\right)$. For multi-way join queries, an EM algorithm with cache complexity $O\left(\frac{N^{\rho^*}}{M^{\rho^*-1}B} \cdot \log_{\frac{M}{B}} \frac{N}{B} + \frac{\OUT}{B}\right)$ has been achieved for Berge-acyclic joins~\cite{hu2016efficient}, $\alpha$-acyclic joins~\cite{hu2021cover, koutris2016worst}, graph joins~\cite{ketsman2017worst,deng2024subgraph} and Loomis-Whitney joins~\cite{koutris2016worst}.\footnote{Some of these algorithms have been developed for the {\em Massively Parallel Computation} (MPC) model~\cite{beame2017communication} and can be adapted to the EM model through the MPC-to-EM reduction~\cite{koutris2016worst}.} These results were previously stated without including the output-dependent term $\frac{\OUT}{B}$ since they do not consider the cost of writing join results back to disk. Again, even padding the output size to be as large as the worst case, these algorithms remain non-oblivious since their access patterns heavily depend on the input data. Furthermore, even in the insecure setting, no algorithm with a cache complexity $O\left(\frac{N^{\rho^*}}{B}\right)$ is known for general join queries. 

    \paragraph{Oblivious Join Algorithms.} Oblivious algorithms have been studied for join queries in both the RAM and EM models. In the RAM model, the naive nested-loop algorithm can be transformed into an oblivious one by incorporating some dummy writes, as it enumerates all possible combinations of tuples from input relations in a fixed order. This algorithm runs in $O(N^{|\E|})$ time, where $|\E|$ is the number of relations in the join query. Wang et al.~\cite{wang2022query} designed circuits for conjunctive queries - capturing all join queries as a special case - whose time complexity matches the AGM bound up to poly-logarithmic factors. Running such a circuit will automatically yield an oblivious join algorithm with $O\left(N^{\rho^*} \cdot \mathsf{polylog} N\right)$ time complexity. By integrating the insecure \wcoj~algorithm~\cite{ngo2014skew} with the optimal ORAM~\cite{asharov2020optorama}, it is possible to achieve an oblivious algorithm with $O(N^{\rho^*} \cdot \log N)$ time complexity, albeit under restrictive theoretical assumptions. Alternatively, incorporating the insecure \wcoj~algorithm into the Path ORAM yields an oblivious join algorithm with $O\left(N^{\rho^*} \cdot \log^2 N \right)$ time complexity. 

    In the EM model, He et al. \cite{he2006cache} proposed a cache-agnostic nested-loop join algorithm for the basic two-way join $R \Join S$ with $O\left(\frac{|R| \cdot |S|}{B}\right)$ cache complexity, which is also oblivious. Applying worst-case padding and the optimal ORAM construction to the existing EM join algorithms, we can derive an oblivious join algorithm with $O\left(\frac{N^{\rho^*}}{B} \cdot \log_{\frac{M}{B}} \frac{N}{B} \cdot \log N\right)$ cache complexity for specific cases such as acyclic joins, graph joins and Loomis-Whitney joins. However, these algorithms are not cache-agnostic. For general join queries, no specific oblivious algorithm has been proposed for the EM model, aside from results derived from the oblivious RAM join algorithm. These results yield cache complexities of either $O\left(N^{\rho^*} \cdot \log N\right)$ or $O\left(N^{\rho^*} \cdot \log N \cdot \log_B N \right)$, as they rely heavily on retrieving tuples from hash tables or range search indices.

    \begin{table}[t]
    \centering
    \begin{tabular}{c|c|c}
        \toprule
         & Previous Results & New Results  \\
         \hline
        \multirow{3}{*}{RAM model} & \multirow{3}{*}{$O\left(N^{\rho^*} \cdot \log N\right)$ \cite{ngo2014skew,asharov2020optorama}} & \multirow{2}{*}{{\color{red} $O\left(N^{\rho^*} \cdot \log N\right)$}} \\
         & \multirow{3}{*}{(one-way function assumption)} & \multirow{2}{*}{\color{red}(no assumption)}\\ 
         & & \\ \cline{1-1} \cline{3-3}
        \multirow{3}{*}{Cache-agnostic} & \multirow{4}{*}{{\color{red} $O\left(\frac{N^{\min\{\rho^*+1, \rho\}}}{B} \cdot \log_{\frac{M}{B}} \frac{N^{\min\{\rho^*+1, \rho\}}}{B} \right)$}}  & \multirow{3}{*}{{\color{red}$O\left(\frac{N^{\rho^*}}{B} \cdot \log_{\frac{M}{B}} \frac{N^{\rho^*}}{B}\right)$}} \\ \cline{2-2}
        \multirow{3}{*}{EM model}& & \\
        
        & \multirow{2}{*}{\color{red}(no assumption)} & \multirow{2}{*}{{\color{red} (tall cache and wide block assumptions)}}\\
        &  &  \\
        \bottomrule
    \end{tabular}
    \caption{Comparison between previous and new oblivious algorithms for multi-way joins. $N$ is the input size. $\rho^*$ and $\rho$ are the input join query's fractional and integral edge cover numbers, respectively. $M$ is the trusted memory size. $B$ is the block size.}
    \label{tab:summary}
\end{table}

    \paragraph{Relaxed Variants of Oblivious Join Algorithms.} 
    Beyond fully oblivious algorithms, researchers have explored relaxed notions of obliviousness by allowing specific types of leakage, such as the join size, the multiplicity of join values, and the size of intermediate results. One relevant line of work examines join processing with released input and output sizes. For example, integrating an insecure output-sensitive join algorithm into an ORAM framework produces a relaxed oblivious algorithm with $O\left((N^{\textsf{subw}}+\OUT) \cdot \mathrm{polylog} N\right)$ time complexity. It is noted that relaxed oblivious algorithm with the same time complexity $O((N+\OUT) \cdot \log N)$ have been proposed without requiring ORAM~\cite{arasu2013oblivious, krastnikov2020efficient} for the basic two-way join as well as acyclic joins. Although not fully oblivious, these algorithms serve as fundamental building blocks for developing our oblivious algorithms for general join queries. Another line of work considered {\em differentially oblivious}  algorithms~\cite{chan2019foundations, beimel2019exploring, chu2021differentially},  which require only that access patterns appear similar across {\em neighboring} input instances. However, differentially oblivious algorithms have so far been limited to the basic two-way join~\cite{chu2021differentially}. This paper does not pursue this direction further.

\subsection{Our Contribution}
Our main contribution can be summarized as follows (see Table~\ref{tab:summary}):
\begin{itemize}
    \item We give a nested-loop-based algorithm for general join queries with $O\left(N^{\min\{\rho^*+1, \rho\}} \cdot \log N\right)$ time complexity and $O\left(\frac{N^{\min\{\rho^*+1, \rho\}}}{B} \cdot \log_{\frac{M}{B}} \frac{N^{\min\{\rho^*+1, \rho\}}}{B}\right)$ cache complexity, where $\rho^*$ and $\rho$ are the fractional and integral edge cover number of the join query, respectively (formally defined in Section~\ref{sec:agm-preliminary}). This algorithm is also cache-agnostic. For classes of join queries with $\rho^*=\rho$, such as acyclic joins, even-length cycle joins and boat joins (see Section~\ref{sec:nested-loop}), this is almost optimal up to logarithmic factors. 
    \item We design an oblivious algorithm for general join queries with $O\left(N^{\rho^*} \cdot \log N\right)$ time complexity, which has matched the insecure counterpart by a logarithmic factor and recovered the previous ORAM-based result, which assumes the existence of one-way functions. 
    This algorithm is also cache-agnostic, with $O\left(\frac{N^{\rho^*}}{B} \cdot \log_{\frac{M}{B}} \frac{N^{\rho^*}}{B} \right)$ cache complexity.
    This cache complexity can be simplified to $O\left(\frac{N^{\rho^*}}{B} \cdot \log_{\frac{M}{B}} \frac{N}{B} \right)$ when $B < N^{\frac{c-\rho^*}{c-1}}$ for some sufficiently large constant $c$. This result establishes the first worst-case near-optimal join algorithm in the insecure EM model when all join results are returned to disk.
\item We develop an improved algorithm for relaxed two-way joins with better cache complexity, which is also cache-agnostic. By integrating our oblivious algorithm with generalized hypertree decomposition~\cite{gottlob2002hypertree}, we obtain a relaxed oblivious algorithm for general join queries with $O\left((N^\textsf{fhtw} + \OUT) \cdot \log N \right)$ time complexity and $O\left(\frac{N^\textsf{fhtw} + \OUT}{B} \cdot \log_{\frac{M}{B}} \frac{N^\textsf{fhtw} + \OUT}{B}\right)$ cache complexity, where $\textsf{fhtw}$ is the fractional hypertree width of the input query.
\end{itemize}

\noindent {\bf Roadmap.} This paper is organized as follows. In Section~\ref{sec:preliminary}, we introduce the preliminaries for building our algorithms. In Section~\ref{sec:nested-loop}, we show our first algorithm based on the nested-loop algorithm. While effective, this algorithm is not always optimal, as demonstrated with the triangle join. In Section~\ref{sec:triangle}, we use triangle join to demonstrate the leakage of insecure \wcoj~algorithm and show how to transform it into an oblivious algorithm. We introduce our oblivious \wcoj~algorithm for general join queries in Section~\ref{sec:general}, and conclude in Section~\ref{sec:conclusion}. 

\section{Preliminaries}
\label{sec:preliminary}
\subsection{Fractional and Integral Edge Cover Number}
 \label{sec:agm-preliminary}
 For a join query $\Q = (\V,\E)$, a function $W: \E \to [0,1]$ is a {\em fractional edge cover}  for $\Q$ if $\sum_{e \in \E:  x \in e} W(e) \ge 1$ for any $x \in \V$. 
 An {\em optimal fractional edge cover} is the one minimizing $\sum_{e \in \E} W(e)$, which is captured by the following linear program: 
\begin{align}
\label{eq:edge-cover}
\min \ \sum_{e \in \E} W(e) \ \ \
    \textrm{s.t. } \ \sum_{e \in \E: x \in e} W(e)& \ge 1, \forall x \in \V; \ \ 
    W(e) \in [0,1], \forall e \in \E 
\end{align}
The optimal value of (\ref{eq:edge-cover}) is the {\em fractional edge cover number} of $\Q$, denoted as $\rho^*(\Q)$. Similarly, a function $W: \E \to \{0,1\}$ is an {\em integral edge cover} if $\sum_{e \in \E:  x \in e} W(e) \ge 1$ for any $x \in \V$. The {\em optimal integral edge cover} is the one minimizing $\sum_{e \in \E} W(e)$, which can be captured by a similar linear program as (\ref{eq:edge-cover}) except that $W(e) \in [0,1]$ is replaced with $W(e) \in \{0,1\}$. The optimal value of this linear program is the {\em integral edge cover number} of $\Q$, denoted as $\rho(\Q)$. 

\subsection{Oblivious Primitives}
\label{sec:oblivious-primitives}
    We introduce the following oblivious primitives, which form the foundation of our algorithms. Each primitive displays an identical access pattern across instances of the same input size. 

    \paragraph{Linear Scan.} Given an input array of $N$ elements, a linear scan of all elements can be done with $O(N)$ time complexity and $O(\frac{N}{B})$ cache complexity in a cache-agnostic way. 
  
    \paragraph{\textsc{Sort}~\cite{ajtai19830, batcher1968sorting}.} Given an input array of $N$ elements, the goal is to output an array according to some predetermined ordering. The classical bitonic sorting network~\cite{batcher1968sorting} requires $O(N \cdot \log^2 N)$ time. Later, this time complexity has been improved to $O\left(N \cdot \log N\right)$~\cite{ajtai19830} in 1983. However, due to the large constant parameter hidden behind $O(N \cdot \log N)$, the classical bitonic sorting is more commonly used in practice, particularly when the size $N$ is not too large. Ramachandran and Shi \cite{ramachandran2021data} showed a randomized algorithm for sorting with $O(N \cdot \log N)$ time complexity and $O\left(\frac{N}{B} \log_{\frac{M}{B}} \frac{N}{B}\right)$ cache complexity under the tall cache assumption. 

    \paragraph{\textsc{Compact}~\cite{ goodrich2011data,  sasy2022fast}.} Given an input array of $N$ elements, some of which are distinguished as $\perp$, the goal is to output an array with all non-distinguished elements moved to the front before any $\perp$, while preserving the ordering of non-distinguished elements.
    Lin et al.~\cite{lin2019can} showed a randomized algorithm for compaction with $O(N \cdot \log \log N)$ time complexity and $O\left(\frac{N}{B}\right)$ cache complexity under the tall cache assumption. 

    \smallskip

    We use the above primitives to construct additional building blocks for our algorithms. To ensure obliviousness, all outputs from these primitives include a fixed size equal to the worst-case scenario, i.e., $N$, comprising both real and dummy elements. All these primitives achieve $O(N \cdot \log N)$ time complexity and $O\left(\frac{N}{B} \cdot \log_{\frac{M}{B}} \frac{N}{B}\right)$ cache complexity. Further details are provided in Appendix~\ref{appendix:pseudocode-primitives}.
    
    \paragraph{\textsc{SemiJoin}.} Given two input relations $R$, $S$ of at most $N$ tuples and their common attribute(s) $x$, the goal is to output the set of tuples in $R$ that can be joined with at least one tuple in $S$.     

    \paragraph{\textsc{Project}.} Given an input relation $R$ of $N$ tuples defined over attributes $e$, and a subset of attribute{s} $x \subseteq e$, the goal is to output $\{t \in R: \pi_{x} t\}$, ensuring no duplication.
  
    \paragraph{\textsc{Intersect}.} Given two input arrays $R,S$ of at most $N$ elements, the goal is to output $R \cap S$. 

    \paragraph{\textsc{Augment}.} Given a relation $R$ and $k$ additional relations $S_1,S_2,\cdots,S_k$ (each with at most $N$ tuples) sharing common attribute(s) $x$, the goal is to attach each tuple $t$ the number of tuples in $S_i$ (for each $i \in [k]$) that can be joined with $t$ on $x$.
    
    \smallskip
    We note that any sequential composition of oblivious primitives yields more complex algorithms that remain oblivious, which is the key principle underlying our approach.

    \subsection{Oblivious Two-way Join}
    \label{sec:binary-join}
    \noindent {\bf \textsc{NestedLoop}.} Nested-loop algorithm can compute $R \Join S$ with $O(|R| \cdot |S|)$ time complexity, which iterates all combinations of tuples from $R, S$ and writes a join result (or a dummy result, if necessary, to maintain obliviousness). He et al. \cite{he2006cache} proposed a cache-agnostic version in the EM model with $O\left(\frac{|R| \cdot |S|}{B}\right)$ cache complexity, which is also oblivious. 

    \begin{theorem}[\cite{he2006cache}]
    \label{the:relaxed}
    For $R \Join S$, there is a cache-agnostic algorithm that
    can compute $R \Join S$ with $O\left(|R| 
    \cdot |S|\right)$ time complexity and $O\left(\frac{|R| \cdot |S|}{B}\right)$ cache complexity, whose access pattern only depends on $M,B, |R|$ and $|S|$.
    \end{theorem}
    \noindent {\bf RelaxedTwoWay.} The relaxed two-way join algorithm~\cite{arasu2013oblivious,krastnikov2020efficient} takes as input two relations $R, S$ and a parameter $\tau \ge |R\Join S|$, and output a table of $\tau$ elements containing join results of $R \Join S$, whose access pattern only depends on $|R|, |S|$ and $\tau$. This algorithm can also be easily transformed into a cache-agnostic version with $O((|R|+|S| +\tau) \cdot \log (|R| + |S| +\tau))$ time complexity and $O\left(\frac{|R|+|S| +\tau}{B} \cdot \log \tau \right)$ cache complexity. In Appendix~\ref{appendix:relaxed}, we show how to improve this algorithm with better cache complexity without sacrificing the time complexity.

    \begin{theorem}
    \label{the:relaxed}
    For $R \Join S$ and a parameter $\tau \ge |R \Join S|$, there is a cache-agnostic algorithm that
    can compute $R \Join S$ with $O\left((|R| 
    +|S| + \tau) \cdot \log (|R|+ |S|+ \tau)\right)$ time complexity and $O\left(\frac{|R| 
    +|S|+ \tau}{B} \cdot \log_{\frac{M}{B}} \frac{|R|+ |S|+ \tau}{B}\right)$ cache complexity under the tall cache and wide block assumptions, whose access pattern only depends on $M, B, |R|,|S|$ and $\tau$.
    \end{theorem}

    \section{Beyond Oblivious Nested-loop Join}
    \label{sec:nested-loop}
    Although the nested-loop join algorithm is described for the two-way join, it can be extended to multi-way joins. For a general join query with $k$ relations, the nested-loop primitive can be recursively invoked $k-1$ times, resulting in an oblivious algorithm with $O\left(\frac{N^k}{B}\right)$ cache complexity. Careful inspection reveals that we do not necessarily feed all input relations into the nested loop; instead, we can restrict enumeration to combinations of tuples from relations included in an integral edge cover of the join query. Recall that for $\Q = (\V,\E)$, an integral edge cover of $\Q$ is a function $W: \E \to \{0,1\}$, such that $\sum_{e: x\in e} W(e) \ge 1$ holds for every $x \in \V$. While enumerating combinations of tuples from relations ``chosen'' by $W$, we can apply semi-joins using remaining relations to filter intermediate join results.  

    As described in Algorithm~\ref{alg:nestedloop}, it first chooses an optimal integral edge cover $W^*$ of $\Q$ (line 1), and then invokes the {\sc NestedLoop} primitive to iteratively compute the combinations of tuples from relations with $W^*(e) = 1$ (line 7), whose output is denoted as $L$. Meanwhile, we apply the semi-join between $L$ and the remaining relations (line 8). 

    Below, we analyze the complexity of this algorithm. First, as $|\E'| \le \rho$, the intermediate join results materialized in the while-loop is at most $O(N^{\rho})$. After semi-join filtering, the number of surviving results is at most $O\left(N^{\rho^*}\right)$. In this way, the number of intermediate results materialized by line 7 is at most $O\left(N^{\rho^*+1}\right)$. 
    Putting everything together, we obtain:
    \begin{theorem}
    \label{the:nested-loop}
    For a general join query $\Q$, there is an oblivious and cache-agnostic algorithm that can compute $\Q(\R)$ for an arbitrary instance $\R$ of input size $N$ with $O\left(N^{\min\{\rho, \rho^*+1\}}\right)$ time complexity and $O\left(\frac{N^{\min\{\rho, \rho^*+1\}}}{B} \cdot \log_{\frac{M}{B}} \frac{N^{\min\{\rho, \rho^*+1\}}}{B}\right)$ cache complexity under the tall cache and wide block assumptions, where $\rho^*$ and $\rho$ are the optimal fractional and integral edge cover number of $\Q$, respectively.  
    \end{theorem}

    It is important to note that any oblivious join algorithm incurs a cache complexity of $\Omega\left(\frac{N^{\rho^*}}{B}\right)$, so Theorem~\ref{the:nested-loop} is optimal up to a logarithmic factor for join queries where $\rho = \rho^*$. Below, we list several important classes of join queries that exhibit this desirable property:

    \begin{algorithm}[t]    \caption{\textsc{ObliviousNestedLoopJoin}$(\Q,\R)$}
    \label{alg:nestedloop}

    $W^* \gets$ an optimal integral edge cover of $\Q$, $L \gets \emptyset$\;
    $\E' \gets \{e \in \E: W^*(e) = 1\}$\; 
    \While{$\E'\neq \emptyset$}{
    $e \gets$ an arbitrary relation in $\E'$\;
    $\E' \gets \E' - \{e\}$\;
    \lIf{$L = \emptyset$}{$L \gets R_e$}
    \lElse{$L \gets \textsc{NestedLoop}(L,R_e)$}
    \lForEach{$e' \in \E-\{e\}$}{$L \gets \textsc{SemiJoin}(L,R_{e'})$}
    }
    \Return $L$\;
    \end{algorithm}

\begin{example}[$\alpha$-acyclic Join]
\label{ex:acyclic}  
A join query $\Q$ is $\alpha$-acyclic~\cite{beeri1983desirability, fagin1983degrees} if there is a tree structure $\mathcal{T}$ of $\Q = (\V,\E)$ such that (1) there is a one-to-one correspondence between relations in $\Q$ and nodes in $\mathcal{T}$; (2) for every attribute $x \in \V$, the set of nodes containing $x$ form a connected subtree of $\mathcal{T}$. Any $\alpha$-acyclic join admits an optimal fractional edge cover that is integral~\cite{hu2021cover}.
\end{example}

\begin{example}[Even-length Cycle Join]
\label{ex:even-cycle}
    An even-length cycle join is defined as $\Q = R_1(x_1,x_2) \Join R_2(x_2,x_3) \Join \cdots \Join R_{k-1}(x_{k-1},x_k)  \Join R_k(x_k,x_1)$ for some even integer $k$. It has two integral edge covers $\{R_1,R_3,\cdots,R_{k-1}\}$ and $\{R_2,R_4,\cdots,R_k\}$, both of which are also an optimal fractional edge cover. Hence, $\rho^* = \rho = \frac{k}{2}$. 
    \end{example}

    \begin{example}[Boat Join] 
    \label{ex:boat}
    A boat join is defined as $\Q = R_1(x_1,y_1) \Join R_2(x_2,y_2) \Join \cdots \Join R_{k}(x_k,y_k) \Join R_{k+1}(x_1,x_2, \cdots, x_k) \Join R_{k+2}(y_1,y_2, \cdots, y_k)$. It has an integral edge cover $\{R_1,R_2\}$ that is also an optimal fractional edge cover. Hence, $\rho^* = \rho = 2$. 
    \end{example}

   \section{Warm Up: Triangle Join}
    \label{sec:triangle}
    The simplest join query that oblivious nested-loop join algorithm cannot solve optimally is the triangle join $\Q_\triangle = R_1(x_2,x_3) \Join R_2(x_1,x_3) \Join R_3(x_1,x_2)$, which has $\rho = 2$ and $\rho^* = \frac{3}{2}$. While various worst-case optimal algorithms for the triangle join have been proposed in the RAM model, none of these are oblivious due to their inherent leakage of intermediate statistics. Below, we outline the issues with existing insecure algorithms and propose a strategy to make them oblivious.

    \paragraph{Insecure Triangle Join Algorithm~\ref{alg:triangle-I}.} We start with attribute $x_1$. Each value $a \in \dom(x_1)$ induces a subquery $\displaystyle{\Q_a = R_1 \Join (\sigma_{x_1 = a}R_2) \Join (\sigma_{x_1 = a}R_3)}$.  Moreover, a value $a \in \dom(x_1)$ is {\em heavy} if $|\pi_{x_3}\sigma_{x_1=a} R_2| \cdot |\pi_{x_2}\sigma_{x_1=a} R_3|$ is greater than $|R_1|$, and {\em light} otherwise. 
    If $a$ is light, $\Q_a$ is computed by materializing the Cartesian product between $\pi_{x_3} \sigma_{x_1 = a} R_1$ and $\pi_{x_2}\sigma_{x_1 = a} R_3$, and then filter the intermediate result by a semi-join with $R_1$. Every surviving tuple forms a join result with $a$, which will be written back to untrusted memory. If $a$ is heavy, $\Q_a$ is computed by applying the semi-joins between $R_1$ with $\sigma_{x_1 = a} R_2$ and $\sigma_{x_1 = a} R_3$. 
    This algorithm achieves a time complexity of  $O(N^{\frac{3}{2}})$ (see \cite{ngo2018worst} for detailed analysis), but it leaks sensitive information through the following mechanisms:
    \begin{itemize}     
        \item $\left|(\pi_{x_1} R_2) \cap (\pi_{x_1} R_3)\right|$ is leaked by the number of for-loop iterations in line 2;
        \item $\left|\pi_{x_2} \sigma_{x_1 = a} R_3\right|$ and $\left|\pi_{x_3} \sigma_{x_1 = a} R_2\right|$ are leaked by the number of for-loop iterations in line 4;
        \item The sizes of heavy and light values in $(\pi_{x_1} R_2) \cap (\pi_{x_1} R_3)$ are leaked by the if-else condition in lines 3 and 6;
    \end{itemize}

   To protect intermediate statistics, we pad dummy tuples to every intermediate result (such as $(\pi_{x_1} R_2) \cap (\pi_{x_1} R_3)$, $\pi_{x_3} \sigma_{x_1 = a} R_2$ and $\pi_{x_2} \sigma_{x_1 = a} R_3$) to match the worst-case size $N$. To hide heavy and light values, we replace conditional if-else branches with a unified execution plan by visiting every possible combination of $\left(\pi_{x_2} \sigma_{x_1 = a} R_3\right) \times \left(\pi_{x_3} \sigma_{x_1 = a} R_2\right)$ and every tuple of $R_1$. By integrating these techniques, this strategy leads to $N^2$ memory accesses, hence destroying the power of two choices that is a key advantage in the insecure \wcoj~algorithm.

   \begin{algorithm}[t]
    \caption{Compute $\Q_\triangle$ by power of two choices}
    \label{alg:triangle-I}
    $L \gets \emptyset$\;
    \ForEach{$a \in \left(\pi_{x_1} R_2\right) \cap \left(\pi_{x_1} R_3\right)$}{
        \If{$|\sigma_{x_1 = a} R_2| \cdot |\sigma_{x_1=a} R_3| \le |R_1|$}{
        \ForEach{$(b,c) \in \left(\pi_{x_2} \sigma_{x_1 = a} R_3\right)\times\left(\pi_{x_3} \sigma_{x_1 = a} R_2\right)$}{
            \lIf{$(b,c) \in R_1$}{\wt $(a,b,c)$ to $L$}  
            }
        }
        \Else{
        \ForEach{$(b,c) \in R_1$}{
                \lIf{$(a,b) \in R_3$ and $(a,c) \in R_2$}{\wt $(a,b,c)$ to $L$}
            }
        }
    }
    \Return $L$\;
    \end{algorithm}
    \begin{algorithm}[t]
    \caption{Inject Obliviousness to Algorithm~\ref{alg:triangle-I}}
    \label{alg:oblivious-triangle-I}
    $A \gets \left(\pi_{x_1} R_2\right) \cap \left(\pi_{x_1} R_3\right)$ by \textsc{Project} and \textsc{Intersect}\;
    $A \gets \textsc{Augment}(A, \{R_2, R_3\}, x_1)$\; 
    $A_1,A_2 \gets \emptyset$\;
    \While(\tcp*[h]{\color{blue}{$\Delta_1 = |\pi_{x_3}\sigma_{x_1 =a}R_2|$ and $\Delta_2 = |\pi_{x_2}\sigma_{x_1=a}R_3|$}}){{\rm \re $(a, \Delta_1, \Delta_2)$ from} $A$}{
        \lIf{$\Delta_1 \cdot \Delta_2 \le |R_1|$}{\wt $a$ to $A_1$, \wt $\perp$ to $A_2$}
        \lElse{\wt $a$ to $A_2$, \wt $\perp$ to $A_1$}
    }
    $L_1 \gets \textsc{RelaxedTwoWay}(A_2, R_1, N^{\frac{3}{2}})$\;
    $L_1 \gets \textsc{SemiJoin}(L_1, R_2)$, $L_1 \gets \textsc{SemiJoin}(L_1, R_3)$\;
    $R_2 \gets \textsc{SemiJoin}(R_2, A_1)$, $R_3 \gets \textsc{SemiJoin}(R_3, A_1)$\;
    $L_2 \gets \textsc{RelaxedTwoWay}(R_2, R_3, N^{\frac{3}{2}})$\;
    $L_2 \gets \textsc{SemiJoin}(L_2, R_1)$\;
    \Return $\textsc{Compact}$ $L_1 \cup L_2$ while keeping the first $N^{\frac{3}{2}}$ tuples\;
    \end{algorithm}

    \paragraph{Inject Obliviousness to Algorithm~\ref{alg:triangle-I}.} To inject obliviousness into Algorithm~\ref{alg:triangle-I}, Algorithm~\ref{alg:oblivious-triangle-I} leverages oblivious primitives to ensure the same access pattern across all instances of the input size. Here's a breakdown of how this is achieved and why it works. We start with computing $A = \left(\pi_{x_1} R_2\right) \cap \left(\pi_{x_1} R_3\right)$ by the {\sc Intersect} primitive. Then, we partition values in $A$ into two subsets $A_1,A_2$, depending on the relative order between 
    $|\pi_{x_3}\sigma_{x_1 =a} R_2| \cdot |\pi_{x_2}\sigma_{x_1 =a}R_3|$ and $|R_1|$.  
    We next compute the following two-way joins $A_2 \Join R_1$ and $(R_2 \ltimes A_1) \Join (R_3 \ltimes A_2)$ by invoking the \textsc{RelaxedTwoWay} primitive separately, each with the upper bound $N^{\frac{3}{2}}$. At last, we filter intermediate join results by the {\sc SemiJoin} primitive and remove unnecessary dummy tuples by the {\sc Compact} primitive.

    \paragraph{Analysis of Algorithm~\ref{alg:oblivious-triangle-I}.} It suffices to show that $|(R_2 \ltimes A_1) \Join (R_3\ltimes A_1)| \le N^{\frac{3}{2}}$ and $|A_2 \Join R_1| \le N^{\frac{3}{2}}$, which directly follows from the query decomposition lemma~\cite{ngo2014skew}:
    \begin{equation*}
        \sum_{a \in A} \min\left\{\left|\sigma_{x_1 =a} R_2\right| \cdot \left|\sigma_{x_1 =a} R_3\right|, |R_1|\right\} 
        \le \sum_{a \in A}  \left(\left|  R_2 \ltimes a \right|  \cdot \left| R_3 \ltimes a \right|\right)^{\frac{1}{2}} \cdot \left| R_1 \ltimes a \right|^{\frac{1}{2}} 
        \le  N^{\frac{3}{2}}.
    \end{equation*}
    All other primitives have $O(N \cdot \log N)$ time complexity and $O\left(\frac{N}{B} \cdot \log_{\frac{M}{B}} \frac{N}{B}\right)$ cache complexity. Hence, this whole algorithm incurs $O\left(N^{\frac{3}{2}} \cdot \log N\right)$ time complexity and $O\left(\frac{N^{\frac{3}{2}}}{B} \cdot \log_{\frac{M}{B}} \frac{N^{\frac{3}{2}}}{B}\right)$ cache complexity. As each step is oblivious, the composition of all these steps is also oblivious.

    \paragraph{Insecure Triangle Join Algorithm~\ref{alg:triangle-II}.} We start with attribute $x_1$. We first compute the candidate values in $x_1$ that appear in some join results, i.e., $(\pi_{x_1} R_2) \cap (\pi_{x_1} R_3)$. For each candidate value $a$, we retrieve the candidate values in $x_2$ that can appear together with $a$ in some join results, i.e., $\left(\pi_{x_2} \sigma_{x_1 =a} R_3\right) \cap \left(\pi_{x_2} R_1\right)$. Furthermore, for each candidate value $b$, we explore the possible values in $x_3$ that can appear together with $(a,b)$ in some join results. More precisely, every value $c$ appears in $\pi_{x_3}\sigma_{x_2=b} R_1$ as well as $\pi_{x_3} \sigma_{x_1 =a} R_2$ forms a triangle with $a,b$. This algorithm runs in $O(N^{\frac{3}{2}})$ time (see \cite{ngo2014skew} for detailed analysis). Similarly, it is not oblivious as the following intermediate statistics may be leaked:
    \begin{itemize}
        \item $\left|(\pi_{x_1} R_2) \cap (\pi_{x_1} R_3)\right|$ is leaked by the number of for-loop iterations in line 2;
        \item $\left|(\pi_{x_2} \sigma_{x_1=a} R_3) \cap (\pi_{x_2} R_1)\right|$ is leaked by the number of for-loop iterations in line 3;
        \item $\left|(\pi_{x_3} \sigma_{x_2=b} R_2) \cap (\pi_{x_3} \pi_{x_1=a} R_2)\right|$ is leaked by the number of for-loop iterations in line 4;
    \end{itemize}

    \begin{algorithm}[t]
    \caption{Compute $\Q_\triangle$ by delaying computation}
    \label{alg:triangle-II}
    $L \gets \emptyset$\;
    \ForEach{$a \in \left(\pi_{x_1} R_2\right) \cap     \left(\pi_{x_1} R_3\right)$}{
    \ForEach{$b \in \left(\pi_{x_2} \sigma_{x_1 =a} R_3\right) \cap \left(\pi_{x_2} R_1\right)$}{
        \ForEach{$c \in \left(\pi_{x_3} \sigma_{x_2 =b} R_1\right) \cap \left(\pi_{x_3} \sigma_{x_1 = a} R_2\right)$}{
                {\wt $(a,b,c)$ to $L$\;}
            }
        }
    }
    \Return $L$\;
    \end{algorithm}
    \begin{algorithm}[t]
    \caption{Inject Obliviousness to Algorithm~\ref{alg:triangle-II}}
    \label{alg:oblivious-triangle-II}
    $R_3 \gets \textsc{Augment}(R_3, R_1, x_2)$, $R_3 \gets \textsc{Augment}(R_3, R_2, x_1)$\;
    $K_1, K_2 \gets \emptyset$\;
    \While(\tcp*[h]{\color{blue}{Suppose $\Delta_i = |R_i \ltimes \{t\}|$}}){{\rm \re $(t,\Delta_1,\Delta_2)$ from} $R_3$}{
        \lIf{$\Delta_1 \le \Delta_2$}{\wt $t$ to $K_1$, \wt $\perp$ to $K_2$}
        \lElse{\wt $t$ to $K_2$, \wt $\perp$ to $K_1$}
    }
    $L_1 \gets \textsc{RelaxedTwoWay}(K_1, R_1, N^{\frac{3}{2}})$, $L_1 \gets \textsc{SemiJoin}(L_1, R_2)$\;
    $L_2 \gets \textsc{RelaxedTwoWay}(K_2, R_2, N^{\frac{3}{2}})$, $L_2 \gets \textsc{SemiJoin}(L_2, R_1)$\;
    \Return \textsc{Compact} $L_1 \cup L_2$ while keeping the first $N^{\frac{3}{2}}$ tuples\;
    \end{algorithm}

    To achieve obliviousness, a straightforward solution is to pad every intermediate result with dummy tuples to match the worst-case size $N$. However, this would result in $N^3$ memory accesses, which is even less efficient than the nested-loop-based algorithm in Section~\ref{sec:nested-loop}.
    
    \paragraph{Inject Obliviousness to Algorithm~\ref{alg:triangle-II}.}
    We transform Algorithm~\ref{alg:triangle-II} into an oblivious version, presented as Algorithm~\ref{alg:oblivious-triangle-II}, by employing oblivious primitives. The first modification merges the first two for-loops (lines 2–3 in Algorithm~\ref{alg:triangle-II}) into one step (line 1 in Algorithm~\ref{alg:oblivious-triangle-II}). This is achieved by applying the semi-joins on $R_3$ using $R_1,R_2$ separately. Then, the third for-loop (line 4 in Algorithm~\ref{alg:triangle-II}) is replaced with a strategy based on the power of two choices. Specifically, for each surviving tuple $(a,b) \in R_3$, we first compute the size of two lists, $\left|\pi_{x_3} \sigma_{x_2=b} R_1 \right|$ and $\left|\pi_{x_3} \sigma_{x_1=a} R_2 \right|$, and put $(a,b)$ into either $K_1$ or $K_2$, based on the relative order between $\left|\pi_{x_3} \sigma_{x_2=b} R_1 \right|$ and $ \left|\pi_{x_3} \sigma_{x_1=a} R_2 \right|$. 
    We next compute the following two-way joins $K_1 \Join R_1$ and $K_2 \Join R_2$ by invoking the \textsc{RelaxedTwoWay} primitive, each with the upper bound $N^{\frac{3}{2}}$ separately. Finally, we filter intermediate join results by the {\sc SemiJoin} primitive and remove unnecessary dummy tuples by the {\sc Compact} primitive.

    \paragraph{Complexity of Algorithm~\ref{alg:oblivious-triangle-II}.} It suffices to show that $|K_1 \Join R_1| \le N^{\frac{3}{2}}$ and $|K_2 \Join R_2|  \le N^{\frac{3}{2}}$, which directly follows from the query decomposition lemma~\cite{ngo2014skew}:
    \begin{equation*}
        \sum_{(a,b) \in R_3} \min\left\{\left|\pi_{x_3} \sigma_{x_2=b} R_1 \right|,  \left|\pi_{x_3} \sigma_{x_1=a} R_2 \right| \right\}
        \le \sum_{(a.b)\in R_3} \left|R_1\ltimes (a,b)\right|^{\frac{1}{2}} \cdot \left| R_2\ltimes (a,b)\right|^{\frac{1}{2}}
        \le N^{\frac{3}{2}}.
    \end{equation*}
    All other primitives incur $O(N \log N)$ time complexity and $O\left(\frac{N}{B} \cdot \log_{\frac{M}{B}} \frac{N}{B}\right)$ cache complexity. 
    Hence, this algorithm incurs $O\left(N^{\frac{3}{2}} \cdot \log N\right)$ time complexity and $O\left(\frac{N^{\frac{3}{2}}}{B} \cdot \log_{\frac{M}{B}} \frac{N^{\frac{3}{2}}}{B}\right)$ cache complexity. As each step is oblivious, the composition of all these steps is also oblivious.

    \begin{theorem}
    \label{the:triangle}
    For triangle join $\Q_\triangle$, there is an oblivious and cache-agnostic algorithm that
    can compute $\Q(\R)$ for any instance $\R$ of input size $N$ with $O\left(N^{\frac{3}{2}} \cdot \log N\right)$ time complexity and $O\left(\frac{N^{\frac{3}{2}}}{B} \cdot \log_{\frac{M}{B}} \frac{N^{\frac{3}{2}}}{B}\right)$ cache complexity under the tall cache and wide block assumptions.
    \end{theorem}

    \section{Oblivious Worst-case Optimal Join Algorithm}
    \label{sec:general}
     \begin{algorithm}[t]
		\caption{$\textsc{GenericJoin}(\Q = (\V,\E), \R)$~\cite{ngo2014skew}}
		\label{alg:generic}
		\lIf{$|\V|=1$}{\Return $\cap_{e \in \E} R_e$ by \textsc{Intersect}}
		$(I,J) \gets $ an arbitrary partition of $\V$\;
		$\Q_I \gets \textsc{GenericJoin}((I, \E[I]), \left\{\pi_{I} R_e: e \in \E\right\})$\;
		\lForEach{$t \in \Q_I$}{
			$\Q_t \gets \textsc{GenericJoin}((J, \E[J]), \left\{\pi_{J} (R_e \ltimes t): e \in \E\right\})$}
		\Return $\bigcup_{t \in \Q_I} \{t\} \times \Q_t$\;
	\end{algorithm}

    In this section, we start with revisiting the insecure \wcoj~algorithm in Section~\ref{sec:generic} and then present our oblivious algorithm in Section~\ref{sec:our-algorithm} and present its analysis in Section~\ref{sec:our-algorithm-analysis}. Subsequently, in Section~\ref{sec:implication}, we explore the implications of our oblivious algorithm for relaxed oblivious algorithms designed for cyclic join queries.

    \subsection{Generic Join Revisited}
    \label{sec:generic}
    In a join query $\Q = (\V,\E)$, for a subset of attributes $S \subseteq \V$, we use $\Q[S] =(S,\E[S])$ to denote the sub-query induced by attributes in $S$, where $\E[S] =\{e \cap S: e \in \E\}$. For each attribute $x \in \V$, we use $\E_x = \{e \in \E: x \in e\}$ to denote the set of relations containing $x$. The insecure \wcoj~algorithm described in~\cite{ngo2014skew} is outlined in Algorithm~\ref{alg:generic}, which takes as input a general join query $\Q= (\V,\E)$ and an instance $\R$. In the base case, when only one attribute exists, it computes the intersection of all relations. For the general case, it partitions the attributes into two disjoint subsets, $I$ and $J$, such that $I \cap J = \emptyset$ and $I \cup J = \V$. The algorithm first computes the sub-query $\Q[I]$, induced by attributes in $I$, whose join result is denoted $\Q_I$. Then, for each tuple $t \in \Q_I$, it recursively invokes the whole algorithm to compute the sub-query $\Q[J]$ induced by attributes in $J$, over tuples that can be joined with $t$. The resulting join result for each tuple $t$ is denoted as $\Q_t$. Finally, it attaches each tuple in $\Q_t$ with $t$, representing the join results in which $t$ participates. The algorithm ultimately returns the union of all join results for tuples in $\Q_I$. However, Algorithm~\ref{alg:generic} exhibits significant leakage of data statistics that violates the obliviousness constraint, for example:
    \begin{itemize}
        \item $\left|\bigcap_{e \in \E} R_e\right|$ is leaked in line 1;
        \item $\left|\pi_I R_e\right|$ for each relation $e \in \E$ is leaked in line 3;
        \item $|\Q_I|$, $\left|\pi_J \left(R_e \ltimes t\right)\right|$, and $\left|\Q_t\right|$ for each tuple $t \in \Q_I$ are leaked in line 4.
    \end{itemize}
    
    More importantly, this algorithm heavily relies on hashing indexes or range search indexes for retrieving tuples, such that the intersection at line 1 can be computed in $O\left(\min_{e\in \E} |R_e|\right)$ time. However, these indexes do not work well in the external memory model since naively extending this algorithm could result in $O\left(N^{\rho^*}\right)$ cache complexity, which is too expensive. Consequently, designing a \wcoj~algorithm that simultaneously maintains cache locality and achieves obliviousness remains a significant challenge.
           \begin{algorithm}[t]
		\caption{$\textsc{ObliviousGenericJoin}(\Q = (\V,\E), \R)$}
		\label{alg:oblivious-generic}
		\lIf{$|\V|=1$}{\Return $\cap_{e \in \E} R_e$ by \textsc{Intersect}}
		$(I,J) \gets $ a partition of $\V$ such that (1) $|J|=1$; or (2) $|J|=2$ (say $J = \{y,z\}$) and $\E_y -\E_z \neq \emptyset$ and $\E_z - \E_y \neq \emptyset$\;
		\lForEach{$e \in \E$}{$S_e \gets \textsc{Project}(R_e, e\cap I)$}
		$\Q_I \gets \textsc{ObliviousGenericJoin}((I, \E[I]), \left\{S_e: e \in \E\right\})$\;
        \If(\tcp*[h]{\color{blue}{Suppose $J  = \{x\}$}}){$|J|=1$}{
			\lForEach{$e \in \E_x$}{
				 $\Q_I \gets \textsc{Augment}(\Q_I, R_e, e\cap I)$}
            $\{Q_I^e\}_{e \in \E_x} \gets \textsc{Partition-I}(\Q_I, \E_x)$\;
			\ForEach{$e \in \E_x$}{
				$L_e \gets \textsc{RelaxedTwoWay}\left(\Q^{e}_I, R_e, N^{\rho^*(\Q)}\right)$\;
				\lFor{$e' \in \E_x - \{e\}$}{$L_e \gets \textsc{SemiJoin}(L_e, R_{e'})$}
			}
            $L \gets \bigcup_{e \in \E_x} L_e$\;
        }
        \Else(\tcp*[h]{\color{blue}{Suppose $J  = \{y,z\}$}}){
				\lForEach{$e \in \E_y \cup \E_z$}{
					$\Q_I \gets \textsc{Augment}(\Q_I, R_e, e\cap I)$}
				%
                $\{\Q_I^{e_1,e_2}\}_{(e_1,e_2)\in (\E_y - \E_z)\times (\E_z - \E_y) }, \{\Q_I^{e_3}\}_{e_3\in \E_x \cap \E_y}  \gets \textsc{Partition-II}(Q_I, \E_y, \E_z)$\;
                \ForEach{$(e_1,e_2) \in (\E_y - \E_z) \times (\E_z - \E_y)$}{
					$L_{e_1,e_2} \gets \textsc{RelaxedTwoWay}\left( \Q^{e_1,e_2}_I, R_{e_1}, N^{\rho^*(\Q)}\right)$\;
                    $L_{e_1,e_2} \gets \textsc{RelaxedTwoWay} \left(L_{e_1,e_2}, R_{e_2}, N^{\rho^*(\Q)}\right)$\;
                    \lForEach{$e \in \E - \{e_1,e_2\}$}{$L_{e_1,e_2} \gets \textsc{SemiJoin}(L_{e_1,e_2}, R_{e})$}
				}
				\ForEach{$e_3 \in \E_y \cap \E_z$}{
					$L_{e_3} \gets \textsc{RelaxedTwoWay}\left( \Q^{e_3}_I, R_{e_3}, N^{\rho^*(\Q)}\right)$\;
                    \lForEach{$e \in \E - \{e_3\}$}{ $L_{e_3} \gets \textsc{SemiJoin} (L_{e_3}, R_{e})$}
				}
                $L \gets \left(\bigcup_{(e_1,e_2) \in (\E_y - \E_z) \times (\E_z - \E_y)} L_{e_1,e_2}\right) \cup \left(\bigcup_{e_3 \in \E_y \cap \E_z} L_{e_3}\right)$\;
			}        
		\Return \textsc{Compact} $L$ while keeping the first $N^{\rho^*(\Q)}$ tuples\;
	\end{algorithm}
    
      \begin{algorithm}[t]
        \caption{$\textsc{Partition-I}(\Q_I, \E_x)$}
	\label{alg:partition-I}
       \lForEach{$e \in \E_x$}{$\Q^{e}_I \gets \emptyset$}
            \While(\tcp*[h]{\color{blue}{Suppose $\Delta_e(t) = |R_e \ltimes \{t\}|$}}){{\rm \re $(t, \{\Delta_e(t)\}_{e\in \E_x})$ from} $\Q_I$}{
				$e' \gets \arg\min_{e \in \E_x} \Delta_e(t)$\; 
				\wt $t$ to $\Q^{e'}_I$ and \wt $\perp$ to $\Q^{e''}_I$ for each $e''\in \E_x-\{e'\}$\;
            }
        \Return $\{Q_I^e\}_{e \in \E_x}$;
    \end{algorithm}
   \begin{algorithm}[t]
        \caption{$\textsc{Partition-II}(\Q_I, \E_y,\E_z)$}
	   \label{alg:partition-II}
        \lForEach{$(e_1,e_2) \in (\E_y - \E_z) \times (\E_z - \E_y)$}{
					$\Q^{e_1,e_2}_I \gets \emptyset$}
        \lForEach{$e_3 \in \E_y \cap \E_z$}{$\Q^{e_3}_I \gets \emptyset$}
		\While(\tcp*[h]{\color{blue}{Suppose $\Delta_e(t) = |R_e \ltimes \{t\}|$}}){{\rm \re $(t,\{\Delta_e(t)\}_{e\in \E_y \cup \E_z})$ from} $\Q_I$}{
					$\displaystyle{e_1,e_2,e_3 \gets \arg\min_{e \in \E_y - \E_z} \Delta_e(t), \arg\min_{e \in \E_z - \E_y} \Delta_e(t), \arg\min_{e \in \E_y \cap \E_z}\Delta_e(t)}$\;
					\If{$\Delta_{e_1}(t) \cdot \Delta_{e_2}(t) \le \Delta_{e_3}(t)$}{
						\wt $t$ to $\Q^{e_1,e_2}_I$\;
                        \lForEach{$(e_1',e_2') \in (\E_y - \E_z) \times (\E_z - \E_y)-\{(e_1,e_2)\}$}{\wt $\perp$ to  $\Q^{e_1',e_2'}_I$}
                        \lForEach{$e_3' \in \E_y \cap \E_z$}{\wt $\perp$ to $\Q^{e_3'}_I$}
                    }
					\Else{
						\wt $t$ to $\Q^{e_3}_I$\;
                        \lForEach{$(e_1',e_2') \in (\E_y - \E_z) \times (\E_z - \E_y)$}{\wt $\perp$ to  $\Q^{e_1',e_2'}_I$}
                        \lForEach{$e_3' \in \E_y \cap \E_z-\{e_3\}$}{\wt $\perp$ to $\Q^{e_3'}_I$}
                    }            
				}
        \Return $\{\Q_I^{e_1,e_2}\}_{(e_1,e_2)\in (\E_y - \E_z)\times (\E_z - \E_y) }, \{\Q_I^{e_3}\}_{e_3\in \E_x \cap \E_y}$\;
    \end{algorithm}

    \subsection{Our Algorithm}
    \label{sec:our-algorithm}
    Now, we extend our oblivious triangle join algorithms from Section~\ref{sec:triangle} to general join queries, as described in Algorithm~\ref{alg:oblivious-generic}. 
    It is built on a recursive framework: 
    
    \noindent {\bf Base Case: $|\V|=1$.} In this case, the join degenerates to the set intersection of all input relations, which can be efficiently computed by the {\sc Intersect} primitive.
    
    \paragraph{General Case: $|\V| > 1$.} In general, we partition $\V$ into two subsets $I$ and $J$, but with the constraint that $|J|=1$ or $|J|=2$ but the two attributes $y,z$ in $J$ must satisfy $\E_y - \E_z \neq \emptyset$ and $\E_z - \E_y \neq \emptyset$. Similar to Algorithm~\ref{alg:generic}, we compute the sub-query $\Q[I]$ by invoking the whole algorithm recursively, whose join result is denoted as $\Q_I$. To prevent the potential leakage, we must be careful about the projection of each relation involved in this subquery, which is computed by the \textsc{Project} primitive. We further distinguish two cases based on $|J|$:

     \paragraph{General Case 1: $|J| = 1$.} Suppose $J = \{x\}$. Recall that for each tuple $t \in \Q_I$, Algorithm~\ref{alg:generic} computes the intersection $\cap_{e \in \E_x}\left(R_e \ltimes t\right)$ on $x$ in the base case. To ensure this step remains oblivious, we must conceal the size of $R_e \ltimes t$. To achieve this, we augment each tuple $t \in \Q_I$ with its {\em degree} in $R_e$, which is defined as $\Delta_e(t) = \left|R_e \ltimes t\right|$, using the \textsc{Augment} primitive. Then, we partition tuples in $\Q_I$ into $|\E_x|$ subsets based on their smallest {\em degree} across all relations in $\E_x$. The details are described in Algorithm~\ref{alg:partition-I}. Let $\Q^e_I \subseteq \Q_I$ denote the set of tuples whose degree is the smallest in $R_e$, i.e., $e = \arg\min_{e' \in \E_x} \Delta_{e'}(t)$ for each $t \in \Q^e_I$. Whenever we write one tuple $t \in \Q_I$ to one subset, we also write a dummy tuple $\perp$ to the other $|\E_x|-1$ subsets. At last, for each $e \in \E_x$, we compute $R_e \Join \Q^e_I$ by invoking the \textsc{RelaxedTwoWay} primitive (line 9), with upper bound $N^{\rho^*}$, and further filter them by remaining relations with semi-joins (line 10). 

     \paragraph{General Case 2: $|J| = 2$.} Suppose $J = \{y,z\}$. 
     Consider an arbitrary tuple $t \in \Q_I$. Algorithm~\ref{alg:generic} computes the residual query $\left\{\bigcap_{e \in \E_y \cap \E_z} (R_e \ltimes t)\right\} \Join \left\{\bigcap_{e \in \E_y - \E_z} (R_e \ltimes t) \right\}\Join \left\{\bigcap_{e \in \E_z - \E_y} (R_e \ltimes t) \right\}$. Like the case above, we first compute its degree in $R_e$ as $\Delta_e(t)$, by the {\sc Augment} primitive. We then partition tuples in $\Q_I$ into $|\E_y \cap \E_z| + |\E_y - \E_z| \cdot |\E_z - \E_y|$ subsets based on their degrees, but more complicated than Case 1. The details are described in Algorithm~\ref{alg:partition-II}. More specifically, for each $e_3 \in \E_y \cap \E_z$, let 
     \begin{align*}
         \Q^{e_3}_I = \biggl\{t \in \Q_I: & \Delta_{e_3}(t) = \min_{e'' \in \E_y \cap \E_z} \Delta_{e''}(t), \Delta_{e_3}(t) < \min_{e\in \E_y-\E_z, e'\in\E_z-\E_y} \Delta_{e}(t) \cdot\Delta_{e'}(t)\biggl\};
     \end{align*}
     and for each pair $(e_1,e_2) \in (\E_y - \E_z) \times (\E_z - \E_y)$, let 
     \begin{align*}
     \label{eq:Q_I-e_1-e_2-1}
         \Q^{e_1,e_2}_I = \biggl\{ t \in \Q_I: & \Delta_{e_1}(t) \cdot \Delta_{e_2}(t) = \min_{e\in \E_y-\E_z,e'\in\E_z-\E_y} \Delta_{e}(t)\cdot \Delta_{e'}(t) \le \min_{e'' \in \E_y \cap \E_z}\Delta_{e''}(t)\biggl\}
     \end{align*}
     For each $(e_1,e_2) \in (\E_y - \E_z) \times (\E_z - \E_y)$, we compute $R_{e_1} \Join R_{e_2} \Join \Q_I^{e_1,e_2}$ by invoking the \textsc{RelaxedTwoWay} primitive iteratively (line 16-17), with the upper bound $N^{\rho^*(\Q)}$, and filter these results by remaining relations with semi-joins (line 18). 
      For each $e_3 \in \E_y \cap \E_z$, we compute $R_{e_3} \Join \Q^{e_3}_I$ by invoking the \textsc{RelaxedTwoWay} primitive (line 20), with the upper bound $N^{\rho^*(\Q)}$, and filter these results by remaining relations with semi-joins (line 21). 


    \subsection{Analysis of Algorithm~\ref{alg:oblivious-generic}} 
    \label{sec:our-algorithm-analysis}
    \noindent {\bf Base Case: $|\V| =1$.} The obliviousness is guaranteed by the {\sc Intersect} primitive. The cache complexity is $O\left(\frac{N}{B} \cdot \log_{\frac{M}{B}} \frac{N}{B}\right)$. In this case, $\rho^* =1$. Hence, Theorem~\ref{the:general} holds.

    \paragraph{General Case: $|\V|>1$.} By hypothesis, the recursive invocation of {\sc ObliviousGenericJoin} at line 4 takes $O\left(N^{\rho^*(\Q)} \cdot \log N\right)$ time and $O\left(\frac{N^{\rho^*}}{B} \cdot \log_{\frac{M}{B}} \frac{N}{B}\right)$ cache complexity,
    since $\rho^*((I,\E[I])) \le \rho^*(\Q)$. We then show the correctness and complexity for all invocations of \textsc{RelaxedTwoWay} primitive.
    Let $\rho^{*}(\cdot)$ be an optimal fractional edge cover of $\Q$. The real size of the two-way join at line 9 can be first rewritten as:
    \begin{equation*}
    \label{eq:1}
        \sum_{e \in \E_x} \left|R_{e} \Join \Q^{e}_I\right|  = \sum_{e \in \E_x} \sum \limits_{t \in \Q^{e}_I} \left|R_{e} \ltimes t\right| = \sum_{e \in \E_x} \sum \limits_{t \in \Q^{e}_I} \min_{e' \in \E_x} \left|R_{e'}\ltimes t\right| \le \sum_{t \in \Q_I} \prod_{e' \in \E_x} \left|R_{e'} \ltimes t\right|^{\rho^{*}(e')} \le N^{\rho^*}
     \end{equation*}
     where the inequalities follow the facts that $\displaystyle{\sum_{e' \in \E_x} \rho^{*}(e') \ge 1}$,  $\bigcup_{r \in \E_x} \Q^e_I = \Q_I$, and the query decomposition lemma~\cite{ngo2014skew}. 
     Hence, $N^{\rho^*(\Q)}$ is valid upper bound for $R_e \Join \Q^e_I$ for each $e \in \E_x$. The real size of the two-way join at lines 18-19 and line 22 can be rewritten as 
    \begin{align}
    \label{eq:2}
        & \sum_{e_1 \in \E_y - \E_z, e_2 \in \E_z - \E_y} \left|  R_{e_1} \Join R_{e_2} \Join \Q^{e_1,e_2}_I  \right| + \sum_{e_3 \in \E_y \cap \E_z} \left| R_{e_3} \Join \Q^{e_3}_I\right| \nonumber\\
        & = \sum_{e_1 \in \E_y-\E_z, e_2 \in \E_z - \E_y} \sum_{t \in \Q^{e_1,e_2}_I} \left|\left(R_{e_1} \ltimes t\right) \Join \left(R_{e_2} \ltimes t\right)\right| + \sum_{e_3 \in \E_y \cap \E_z} \sum_{t \in \Q^{e_3}_I} \left|R_{e_3} \ltimes t\right| \nonumber \\
        & = \sum_{t \in \Q_I} \min\left\{\min_{e_1\in \E_y - \E_z, e_2 \in \E_z -\E_y}|R_{e_1} \ltimes t| \cdot |R_{e_2} \ltimes t|, \min_{e_3\in \E_y \cap \E_z}|R_{e_3} \ltimes t|\right\} 
    \end{align}
    Let $\displaystyle{\rho_1 = \sum_{e \in \E_y - \E_z} \rho^*(e)}$, $\displaystyle{\rho_2 = \sum_{e \in \E_z - \E_y} \rho^*(e)}$ and $\displaystyle{\rho_3 = \sum_{e \in \E_y  \cap \E_z} \rho^*(e)}$. Note $\rho_3 \ge 1-\min\{\rho_1, \rho_2\}$ as $\rho^*(\cdot)$ is a valid fractional edge cover for both $y$ and $z$. For each tuple $t\in \Q_I$, we have
    \begin{align*}
        & \min\left\{\min_{e_1 \in \E_y - \E_z, e_2 \in \E_z - \E_y} \left|R_{e_1} \ltimes t\right| \cdot |R_{e_2} \ltimes t|, \min_{e_3 \in \E_y \cap \E_z} \left|R_{e_3} \ltimes t\right|\right\} \\
        & \le \left(\min_{e_1 \in \E_y - \E_z} \left|R_{e_1} \ltimes t\right|\right)^{\rho_1} \cdot \left(\min_{e_2 \in \E_z - \E_y} |R_{e_2} \ltimes t|\right)^{\rho_2} \cdot \left(\min_{e_3 \in \E_y \cap \E_z} \left|R_{e_3} \ltimes t\right|\right)^{\rho_3} \\ 
        & \le \prod_{e \in \E_y - \E_z} \left|R_e \ltimes t\right|^{\rho^*(e)} \cdot \prod_{e \in \E_z -\E_y} \left|R_e \ltimes t\right|^{\rho^*(e)} \cdot \prod_{e \in \E_y \cap \E_z} \left|R_e \ltimes t\right|^{\rho^*(e)} = \prod_{e \in \E_y \cup \E_z} \left|R_e \ltimes t\right|^{\rho^*(e)}, 
    \end{align*}
    where the first inequality follows from $\min\left\{a,b\right\}\le a^p\cdot b^{1-p}$ for $a,b\ge 0$ and $p\in [0,1]$, and the third inequality follows from $\rho_1, \rho_2 \ge  \min\left\{\rho_1,\rho_2\right\}$. Now, we can further bound (\ref{eq:2}) as
    \[(\ref{eq:2}) \le \sum_{t \in \Q_I} \prod_{e \in \E_y \cup \E_z} \left|R_{e} \ltimes t\right|^{\rho^*(e)}  = \sum_{t\in \Q_I} \prod_{e \in \E_y \cup \E_z} \left|R_{e} \ltimes t\right|^{\rho^*(e)} \le \prod_{e \in \E}|R_e|^{\rho^*(e)} \le N^{\rho^*}\]
    where the second last inequality follows the query decomposition lemma~\cite{ngo2014skew}.


    \begin{theorem}
    \label{the:general}
    For a general join query $\Q$, there is an oblivious and cache-agnostic algorithm that
    can compute $\Q(\R)$ for any instance $\R$ of input size $N$ with $O\left(N^{\rho^*} \cdot\log N\right)$ time complexity and $O\left(\frac{N^{\rho^*}}{B} \cdot \log_{\frac{M}{B}} \frac{N^{\rho^*}}{B}\right)$ cache complexity under the tall cache and wide block assumptions, where $\rho^*$ is the optimal fractional edge cover number of $\Q$. 
    \end{theorem}
      
     \subsection{Implications to Relaxed Oblivious Algorithms}
    \label{sec:implication}

    Our oblivious \wcoj~algorithm can be combined with 
    the generalized hypertree decomposition framework~\cite{gottlob2002hypertree} to develop a relaxed oblivious algorithm for general join queries. 
    
    \begin{definition}[Generalized Hypertree Decomposition (GHD)]
    \label{def:ghd}
    Given a join query $\Q=(\V,\E)$, a GHD of $\Q$ is a pair $(\T, \lambda)$, where $\T$ is a tree as an ordered set of nodes and $\lambda: \T \to 2^{\V}$ is a labeling function which associates to each vertex $u \in \T$ a subset of attributes in $\V$, $\lambda_u$, such that (1) for each $e \in \E$, there is a node $u \in \T$ such that $e \subseteq \lambda_u$; (2) For each $x \in \V$, the set of nodes $\{u \in \T: x \in \lambda_u\}$ forms a connected subtree of $\T$. The fractional hypertree width of $\Q$ is defined as $ \displaystyle{\min_{(\T,\lambda)} \max_{u \in \T} \rho^*\left((\lambda_u, \{e \cap \lambda: e \in \E\})\right)}$.
    \end{definition}
    
    The pseudocode of our algorithm is given in Appendix~\ref{appendix:general}. Suppose we take as input a join query $\Q=(\V,\E)$, an instance $\R$, and an upper bound on the output size $\tau \ge |\Q(\R)|$. Let $(\T,\lambda)$ be an arbitrary GHD of $\Q$. We first invoke Algorithm~\ref{alg:oblivious-generic} to compute the subquery $\Q_u = (\lambda_u, \E_u)$ defined by each node $u \in \mathcal{T}$, where $\E_u = \{e \cap u: e\in \E\}$, and materialize its join result as one relation. We then apply the classic Yannakakis algorithm~\cite{yannakakis1981algorithms} on the materialized relations by invoking the {\sc SemiJoin} primitive for semi-joins and the \textsc{RelaxedTwoWay} primitive for pairwise joins. After removing dangling tuples, the size of each two-way join is upper bound by the size of the final join results and, therefore, $\tau$. This leads to a relaxed oblivious algorithm whose access pattern only depends on $N$ and $\tau$.
   \begin{theorem}
    \label{the:relaxed}
    For a join query $\Q$, an instance $\R$ of input size $N$, and parameter $\tau \ge |\Q(\R)|$, there is a cache-agnostic algorithm that
    can compute $\Q(\R)$ with $O\left((N^{w} + \tau) \cdot \log (N^w +\tau)\right)$ time complexity and $O\left(\frac{N^{w} + \tau}{B} \cdot \log_{\frac{M}{B}} \frac{N^w +\tau}{B}\right)$ cache complexity, whose access pattern only depends on $N$ and $\tau$, where $w$ is the fractional hypertree width of $\Q$.
    \end{theorem}

    \section{Conclusion}
    \label{sec:conclusion}
    This paper has introduced a general framework for oblivious multi-way join processing, achieving near-optimal time and cache complexity. However, several intriguing questions remain open for future exploration:
    \begin{itemize}
        \item {\em Balancing Privacy and Efficiency:} Recent research has investigated improved trade-offs between privacy and efficiency, aiming to overcome the challenges of worst-case scenarios, such as differentially oblivious algorithms~\cite{chan2019foundations}. 
        \item {\em Emit model for EM algorithms.} In the context of EM join algorithms, the {\em emit} model - where join results are directly outputted without writing back to disk - has been considered. It remains open whether oblivious, worst-case optimal join algorithms can be developed without requiring all join results to be written back to disk. 
        \item {\em Communication-oblivious join algorithm for MPC model.} 
        A natural connection exists between the MPC and EM models in join processing. While recent work has explored communication-oblivious algorithms in the MPC model~\cite{chan2020mpc, tao2024parallel}, extending these ideas to multi-way join processing remains an open challenge.   
        \end{itemize}
        
    \bibliographystyle{abbrv}
    \bibliography{paper}

    \appendix
    \section{Missing Materials in Section~\ref{sec:intro}}
    %
    %
    \noindent {\bf Graph Joins.} A join query $\Q=(\V,\E)$ is a graph join if $|e| \le 2$ for each $e \in \E$, i.e., each relation contains at most two attributes.

    \paragraph{Loomis-Whitney Joins.} A join query $\Q = (\V,\E) $ is a Loomis-Whitney join if $\V = \{x_1,x_2,\cdots, x_k\}$ and $\E = \{\V -\{x_i\}: i \in [k]\}$.

    \section{Oblivious Primitives in Section~\ref{sec:preliminary}}
    \label{appendix:pseudocode-primitives}
    We provide the algorithm descriptions and pseudocodes for the oblivious primitives declared in Section~\ref{sec:oblivious-primitives}. For the local variables used in these primitives, $\key$, $\val$, $\mathsf{pos}$ and $\mathsf{cnt}$, we do not need to establish obliviousness for them because they are stored in the trusted memory during the entire execution of the algorithms and the adversaries cannot observe the access pattern to them. But for all the temporal sets with non-constant size, $K$ and $L$, they are stored in the untrusted memory.
  
\paragraph{\textsc{SemiJoin}.}
    Given two input relations $R$, $S$ and their common attribute(s) $x$, the goal is to replace each tuple in $R$ that cannot be joined with any tuple in $S$ with a dummy tuple $\perp$, i.e., compute $R \ltimes S$. As shown in Algorithm~\ref{alg:semi-join}, we first sort all tuples by their join values and break ties by putting $S$-tuples before $R$-tuples if they share the same join value in $x$. We then perform a linear scan, using an additional variable $\key$ to track the largest join value of the previous tuple that is no larger than the join value of the current tuple $t$ visited. More specifically, we distinguish two cases on $t$. Suppose $t \in R$. If $\pi_{x} t = \key$, we just write $t$ to the result array $L$. Otherwise, we write a dummy tuple $\perp$ to $L$. Suppose $t \in S$. We simply write a dummy tuple $\perp$ to $L$ and update $\key$ with $\pi_{x} t$. At last, we compact the elements in $L$ to move all $\perp$ to the last and keep the first $|R|$ tuples in $L$. 
    
    \begin{algorithm}
	\caption{{\sc SemiJoin}($R, S, x$) }
	\label{alg:semi-join}
	$K \gets $ {\sc Sort} $R \cup S$ by attribute(s) $x$, breaking ties by putting $S$-tuples before $R$-tuples when they have the same value in $x$\;
    $\key \gets \perp$, $L \gets \emptyset$\;
    \While{{\rm \re $t$ from} $K$}{
    \If{$t \in R$}{
        \lIf{$t \neq \perp$ {\rm and} $\pi_{x} t = \key$}{\wt $t$ to $L$}
        \lElse{\wt $\perp$ to $L$} 
    }
    \lElse{\wt $\perp$ to $L$\;
            $\key \gets \pi_{x} t$}
     }
    \Return {\sc Compact} $L$ while keeping the first $|R|$ tuples\;
 	\end{algorithm}

\begin{algorithm}[t]
    \caption{{\sc ReduceByKey}($R, x, w(\cdot), \oplus$)   }
    \label{alg:reduce-by-key}
    $K \gets \textrm{\textsc{Sort}}$ $R$ by attribute(s) $x$ with all $\perp$ moved to the last\;
    $\mathsf{key} \gets \perp$, $\val \gets 0$, $L \gets \emptyset$\;
    \While{{\rm \re $t$ from}  $K$}{
        \lIf{$t = \perp$}{
               \wt $\perp$ to $L$}
        \lElseIf{$t \neq \perp$ {\rm and} $\pi_{x} t = \mathsf{key}$}{
            \wt $\perp$ to $L$, $\val \gets \val \oplus w(t)$}
        \lElse{ \wt $(\textsf{key}, \val)$ to $L$, $\val \gets w(t)$, $\textsf{key} \gets \pi_x t$} 
     }
     \wt $(\textsf{key}, \val)$ to $L$\;
     \Return {\sc Compact} $L$ while keeping the first $|R|$ tuples\;
     \end{algorithm}

    \paragraph{\textsc{ReduceByKey}.}
    Given an input relation $R_e$, some of which are distinguished as $\perp$, a set of key attribute(s) $x \subseteq e$, a weight function $w$, and  an aggregate function $\oplus$, the goal is to output the aggregation of each key value, which is defined as the function $\oplus$ over the weights of all tuples with the same key value. This primitive can be used to compute {\em degree information}, i.e., the number of tuples displaying a specific key value in a relation. 

     As shown in Algorithm~\ref{alg:reduce-by-key}, we sort all tuples by their key values (values in attribute(s) $x$) while moving all distinguished tuples to the last of the relation. Then, we perform a linear scan, using an additional variable $\key$ to track the key value of the previous tuple, and $\val$ to track the aggregation over the weights of tuples visited. We distinguish three cases. If $t = \perp$, the remaining tuples in $K$ are all distinguished as $\perp$, implied by the sorting. We write a dummy tuple $\perp$ to $L$ in this case. If $t \neq \perp$ and $\pi_x t =\key$, we simply write a dummy tuple $t$ to $L$, and increase $\val$ by $w(t)$. If $t \neq \perp$ and $\key \neq \pi_x t$, the values of all elements with key $\key$ are already aggregated into $\val$. In this case, we need to write $(\key, \val)$ to $L$ and update $\val$ with $w(t)$, i.e., the value of current tuple, and $\key$ with $\pi_x t$. At last, we compact the tuples in $L$ by moving all $\perp$ to the last and keep the first $|R|$ tuples in $L$ for obliviousness.  

     \begin{algorithm}[t]
     \caption{{\sc Annotate}($R, S, x)$)}
     \label{alg:annotate}
     $K \gets$ \textsc{Sort} $R \cup S$ by attribute(s) $x$ while moving all $\perp$ to the last and breaking ties by putting $S$-tuples before $R$-tuples when they have the same value in $x$\; 
     $\key \gets \perp$, $\val \gets 0$ , $L \gets \emptyset$\;
     \While{{\rm \re $t$ from}  $K$}{
     	\lIf{$t = \perp$}{\wt $\perp$ to $L$}
        \ElseIf{$t \in S$}{
            \wt $\perp$ to $L$, $\val \gets \pi_{\Bar{x}} t$, $\key \gets \pi_x t$}
        \lElseIf{$t \in R$ {\rm and} $\pi_x t = \key$}{\wt $(t, \val)$ to $L$}
        \lElse{\wt $\perp$ to $L$}
     }    
     \Return \textsc{Compact} $L$ while keeping the first $|R|$ tuples\;
     \end{algorithm}

    \paragraph{\textsc{Annotate}.}
    Given an input relation $R$, where each tuple is associated with a key, and a list $S$ of key-value pairs, where each pair is associated with a distinct key, the goal is to attach, for each tuple in $R$, the value of the corresponding distinct pair in $S$ matched by the key. As shown in Algorithm~\ref{alg:annotate}, we first sort all tuples in $R$ and $S$ by their key values in attribute $x$, while moving all $\perp$ to the last of the relation and breaking ties by putting all $S$-tuples before $R$-tuples when they have the same key value. We then perform a linear scan, using another two variables $\key, \val$ to track the $S$-tuple with the largest key but no larger than the key of the current tuple visited. We distinguish the following cases. If $t$ is a $S$-tuple and $t \neq \perp$, we update $\key, \val$ with $t$.  If $t$ is a $R$-tuple and $t \neq \perp$, we attach $\val$ to $t$ by writing $(t,\val)$ to $L$. We write a dummy tuple $\perp$ to $L$ in the remaining cases. Finally, we compact the tuples in $L$ to remove unnecessary dummy tuples.
    \begin{algorithm}[t]
    \caption{{\sc MultiNumber}($R, x$)}
    \label{alg:multi-number}
    $K \gets$ \textsc{Sort} $R$ by attribute(s) $x$\;
    $\key \gets \perp$, $\val \gets 0$, $L \gets \emptyset$\;
    \ForEach{$t \in K$}{
        \lIf{$\pi_x t = \key$}{ $\val \gets \val+1$ }
        \lElse{$\val \gets 1$, \  $\key \gets \pi_x t$}
        \wt $(t,\val)$ to $L$\;
    }
    \Return $L$\;
    \end{algorithm}

    \paragraph{\textsc{MultiNumber}.}
    Given an input relation $R$, each associated with a key attribute(s) $x$, the goal is to attach consecutive numbers $1,2,3,\cdots,$ to tuples with the same key.
    
    As shown in Algorithm~\ref{alg:multi-number}, we first sort all tuples in $R$ by attribute $x$. We then perform a linear scan, using two additional variables $\key, \val$ to track the key of the previous tuples, and the number assigned to the previous tuple. Consider $t$ as the current element visited. If $\pi_x t = \key$, we simply increase $\val$ by $1$. Otherwise, we set $\val$ to $1$ and update $\key$ with $\pi_x t$. In both cases, we assign $\val$ to tuple $t$ and writw $(t,\val)$ to $L$.

    \paragraph{\textsc{Project}.}
    Given an input relation $R$ defined over attributes $e$, and a subset of attribute{s} $x \subseteq e$, the goal is to output the list $\{t \in R: \pi_{x} t\}$ (without duplication). This primitive can be simply solved by sorting by attribute(s) $x$ and then removing duplicates by a linear scan.

    \paragraph{\textsc{Intersect}.}
    Given two input arrays $R,S$ of distinct elements separately, the goal is to output the common elements appearing in both $R$ and $S$.  This primitive can be done with sorting by attribute(s) $x$, and then a linear scan would suffice to find out common elements.

    \paragraph{\textsc{Augment}.} Given two relations $R,S$ of at most $N$ tuples and their common attribute(s) $x$, the goal is to attach each tuple $t$ the number of tuples in $S$ that can be joined with $t$ on $x$. The {\sc Augment} primitive can be implemented by the {\sc ReduceByKey} and {\sc Annotate} primitives. See Algorithm~\ref{alg:augment}.
    \begin{algorithm}[t]
    \caption{{\sc Augment}($R, \{S_1,S_2,\cdots,S_k\}, x$)}
    \label{alg:augment}
    \ForEach{$i \in [k]$}{
    $L \gets \textsc{ReduceByKey}(S_i, x)$\;
    $R \gets \textsc{Annotate}(R, L, x)$\;
    }
    \Return $R$\;
    \end{algorithm}

    \section{\textsc{RelaxedTwoWay} Primitive}
    \label{appendix:relaxed}
     Given two relations $R, S$ of $N_1, N_2$ tuples and an integral parameter $\tau$, where $N_1+N_2=N$ and $|R \Join S|\le \tau$, the goal is to output a relation of size $\tau$ whose first $|R \Join S|$ tuples are the join results and the remaining tuples are dummy tuples. Arasu et al. \cite{arasu2013oblivious} first proposed an oblivious algorithm for $\tau = |R \Join S|$, but it involves rather complicated primitive without giving complete details~\cite{chang2021efficient}. Krastnikov et al.~\cite{krastnikov2020efficient} later showed a more clean and effective version, but this algorithm does not have a satisfactory cache complexity. Below, we present our own version of the relaxed two-way join. We need one important helper primitive first.

    \paragraph{\textsc{Expand} Primitive.}
    Given a sequence of $\langle (t_i,w_i): w_i \in \mathbb{Z}^+, i \in [N]\rangle$ and a parameter $\tau \ge \sum_{i \in [N]} w_i$, the goal is to expand each tuple $t_i$ with $w_i$ copies and output a table of $\tau$ tuples. The naive way of reading a pair $(t_i,w_i)$ and then writing $w_i$ copies does not preserve obliviousness since the number of consecutive writes can leak the information. Alternatively, one might consider writing a fixed number of tuples after reading each pair. Still, the ordering of reading pairs is critical for avoiding dummy writes and avoiding too many pairs stored in trusted memory (this strategy is exactly adopted by \cite{arasu2013oblivious}). 
    
    We present a simpler algorithm by combining the oblivious primitives. Suppose $L$ is the output table of $R$, such that $L$ contains $w_i$ copies of $t_i$, and any tuple $t_i$ comes before $t_j$ if $i < j$. As described in Algorithm~\ref{alg:FO-expand}, it consists of four phases:
    \begin{itemize}
    \item {\bf (lines 1-4).} for each pair $(t_i, w_i) \in R$ with $w_i \neq 0$, attach the beginning position of $t_i$ in $\Tilde{R}$, which is $\sum_{j < i} w_j$. For the remaining pairs with $w_i = 0$, replace them with $\perp$ and attach with the infinite position as these tuples will not participate in any join result;
    \item {\bf (lines 5-7)} pad $\tau$ dummy tuples and attach them with consecutive numbers $1.5,2.5,\cdots$; after sorting the well-defined positions, each tuple $t_i$ will be followed by $w_i$ dummy tuples, and all dummy tuples with infinite positions are put at last; 
    \item {\bf (lines 8-14)} for each tuple $t_i$, we replace it with $\perp$ but the following $w_i$ dummy tuples with $t_i$.  After moving all dummy tuples to the end, the first $\tau$ elements are the output.
    \end{itemize}

    \begin{algorithm}[t]
    \caption{{\sc Expand}($R=\langle (t_i, w_i):i\in[N]\rangle, \tau$)}
    \label{alg:FO-expand}
    $\mathsf{pos} \gets 1$, $K \gets \emptyset$\;
    \While{{\rm \re $(t_i,w_i)$ from} $R$}{
    \lIf{$(t_i, w_i)=(\perp, \perp)$}{\wt $(\perp, +\infty)$ to $X$}
    \lElse{\wt $(t_i, \mathsf{pos})$ to $K$, $\mathsf{pos} \gets \mathsf{pos} + w_i$}
    } 
    $\mathsf{pos} \gets 1.5$\;
    \lForEach{$i \in [\tau]$}{\wt $(\perp, \mathsf{pos})$ to $K$, $\mathsf{pos} \gets  \mathsf{pos}+1$}
    \textsc{Sort} $K$ by $\mathsf{pos}$\;
    $t \gets \perp$, $\mathsf{cnt} \gets 0$, $L\gets \emptyset$\;
    \While{{\rm \re $(\key, \mathsf{pos})$ from} $K$}{
    \lIf{$\mathsf{pos} = +\infty$}{ \wt $\perp$ to $L$}
    \lElseIf{$\key \neq \perp$}{$t \gets \key$, \wt $\perp$ to $L$}
    \lElseIf{$\mathsf{cnt} < \tau$}{\wt $t$ to $L$ }
    \lElse{\wt $\perp$ to $L$}
    $\mathsf{cnt} \gets \mathsf{cnt} + 1$\;
    }
    \Return  \textsc{Compact} $L$ while keeping the first $\tau$ elements\;
    \end{algorithm}
 
     It can be easily checked that the access pattern of \textsc{Expand} only depends on the values of $\tau$ and $N$. Moreover, \textsc{Expand} is cache-agnostic since they are constructed by sequential compositions of cache-agnostic primitives ({\sc Scan}, {\sc Sort} and {\sc Compact}). 

    \begin{lemma}
    \label{the:expand}
    Given a relation $\R$ of input size $N$ and a parameter $\tau$, the {\sc Expand} primitive is cache-agnostic with $O\left((N + \tau) \cdot \log (N + \tau)\right)$ time complexity and $O\left(\frac{N + \tau}{B} \log_{\frac{M}{B}} \frac{N+\tau}{B}\right)$ cache complexity, whose access pattern only depends on $N$ and $\tau$. 
    \end{lemma}

\begin{sidewaysfigure}
     \centering
     \includegraphics[scale=1.4]{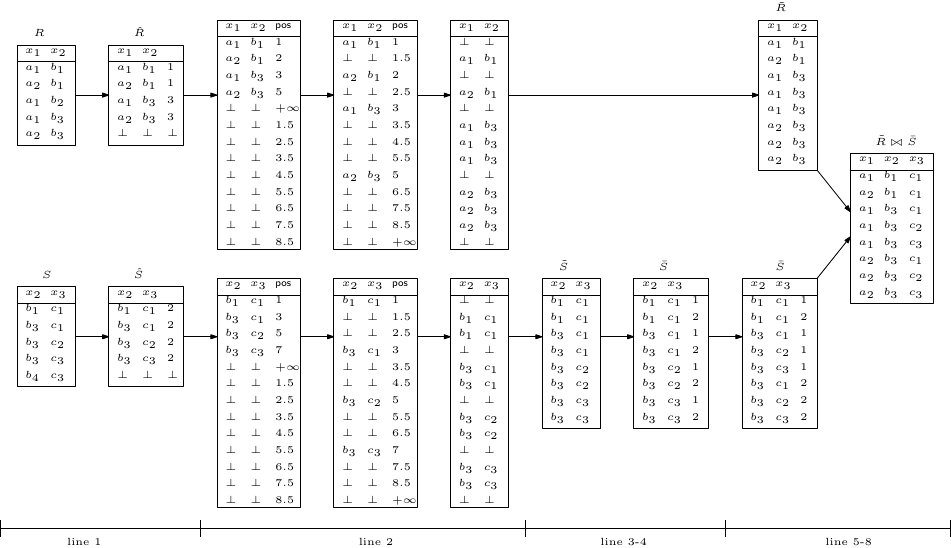}
     \caption{A running example of Algorithm~\ref{alg:relaxed}.}
     \label{fig:relaxed}
     \end{sidewaysfigure}

    Now, we are ready to describe the algorithmic details of {\sc RelaxedTwoWay} primitive. The high-level idea is to simulate the sort-merge join algorithm without revealing the movement of pointers in the merge phase. Let $L = R(x_1,x_2) \Join S(x_2,x_3)$ be the join results sorted by $x_2,x_3,x_1$ lexicographically. The idea is to transform $R, S$ into a sub-relation of $L$ by keeping attributes $(x_1, x_2), (x_2,x_3)$ separately, without removing duplicates. Then, doing a one-to-one merge to obtain the final join results suffices. As described in Algorithm~\ref{alg:relaxed}, we construct these two sub-relations from the input relations $R, S$ via the following steps (a running example is given in Figure~\ref{fig:relaxed}): 
    \begin{itemize}
     \item {\bf (line 1)} attach each tuple with the number of tuples it can be joined in the other relation;
     \item {\bf (line 2)} expand each tuple to the annotated number of copies;
     \item {\bf (lines 3-4)} prepare the expanded $\Tilde{R}$ and $\Tilde{S}$ with the ``correct'' ordering, as it appears in the final sort-merge join results;
      \item {\bf (lines 5-8)} perform a one-to-one merge of ordered tuples in $\Tilde{R}$ and $\Bar{S}$;
    \end{itemize}

    As a sequential composition of (relaxed) oblivious primitives, \textsc{RelaxedTwoWay} is cache-agnostic, with $O((N+\tau) \cdot \log (N+\tau))$ time complexity and $O(\frac{N+\tau}{B} \cdot \log_{\frac{M}{B}} \frac{N+\tau}{B})$ cache complexity, whose access pattern only depends on $N$ and $\tau$.  

    \begin{algorithm}[t]
        \caption{{\sc RelaxedTwoWay}($R(x_1,x_2), S(x_2,x_3), \tau$)}
        \label{alg:relaxed}
        $\hat{R} \gets \textsc{Augment}(R, S, x_2)$, $\hat{S} \gets \textsc{Augment}(S, R, x_2)$\;
        $\Tilde{R} \gets \textsc{Expand} (\hat{R}, \tau)$, $\Tilde{S} \gets \textsc{Expand}(\hat{S}, \tau)$\;
        $\Bar{S} \gets \textsc{MultiNumber}(\Tilde{S}, x_2)$; \tcp{\color{blue}{$\tilde{S}$ is enriched with another attribute $\mathsf{num}$}}
        $\textsc{Sort}$ $\Bar{S}$ by attributes $x_2$ and $\mathsf{num}$ lexicographically\;
        $L \gets \emptyset$\;
         \While{\rm \re $t_1$ from $\Tilde{R}$ and \re $t_2$ from $\Bar{S}$}{
            \lIf{\rm $t_1 \not= \perp$ and $t_2\not=\perp$}{\wt $t_1 \Join t_2$ to $L$}
            \lElse{\wt $\perp$ to $L$}
        }
    \Return $L$\;        
    \end{algorithm}

    \section{Missing Materials in Section~\ref{sec:general}}
    \label{appendix:general}
       \begin{algorithm}[h]
    \caption{$\textsc{RelaxedJoin}(\Q = (\V,\E), \R, \tau)$}
     \label{alg:cyclic-relaxed}
     $(\T, \lambda) \gets$ a GHD of $\Q$\;
     \ForEach{node $u \in \T$}{
        $\E_u \gets \{e \cap \lambda_u: e \in \E\}$\;
        \lForEach{$e \in \E$}{$S_{e,u} \gets \pi_{e \cap \lambda_u} R_e$ by {\sc project}}
        $\Q_u \gets \textsc{OblivousGenericJoin}\left((\lambda_u, \E_u), \{S_{e,u}: e \in \E\}\right)$\;
     }
     \While{visit nodes $u \in \T$ in a bottom-up way (excluding the root)}{
        $p_u \gets $ the parent node of $u$\;
        $\Q_{p_u} \gets \textsc{SemiJoin}(\Q_{p_u}, \Q_u)$\;
     }
     \While{visit nodes $u \in \T$ in a top-down way (excluding the leaves)}{
        \lForEach{child node $v$ of $u$}{
        $\Q_v \gets \textsc{SemiJoin}(\Q_v, \Q_u)$}
    }
    \While{visit nodes $u \in \T$ in a bottom-up way (excluding the root)}{
        $p_u \gets
        $ the parent node of $u$\;
        $\Q_{p_u} \gets \textsc{RelaxedTwoWay}(\Q_{p_u}, \Q_u, \tau)$\;
     }
    \Return $\Q_r$ for the root node $r$ of $\T$\;
    \end{algorithm}

\end{document}